\documentclass[twocolumn,showpacs,preprintnumbers,amsmath,amssymb]{revtex4}

\usepackage{graphicx}
\usepackage{dcolumn}
\usepackage{bm}

\begin{document}


\title{Essential conditions for evolution of communication within a species}

\author{Alex Feigel}
 \email{sasha@soreq.gov.il}
\affiliation{%
Soreq NRC,\\
Yavne 81800, Israel}


\date{\today}

\begin{abstract}
A major obstacle in analyzing the evolution of information exchange and processing is our insufficient understanding of the underlying signaling and decision-making biological mechanisms. For instance, it is unclear why are humans unique in developing such extensive communication abilities. To treat this problem, a method based on the mutual information approach is developed that evaluates the information content of communication between interacting individuals through correlations of their behavior patterns (rather than calculating the information load of exchanged discrete signals, e.g. Shannon entropy). It predicts that correlated interactions of the indirect reciprocity type together with affective behavior and selection rules changing with time are necessary conditions for the emergence of significant information exchange. Population size variations accelerate this development. These results are supported by evidence of demographic bottlenecks, distinguishing human from other species' (e.g. apes) evolution line. They indicate as well new pathways for evolution of information based phenomena, such as intelligence and complexity.
\end{abstract}

\maketitle

Information is a resource, like energy or food, contributing to the evolutionary success of a population and its members. As a consequence, the majority of biological species, from bacteria and plants to human beings, have developed skills to sense their environment, communicate with each other \citep{Hauser2002,Keller2006,Baldwin2006,Dudareva2006} and in some cases even interfere with the signalling of hostile organisms \citep{Bauer2002}. The abilities to acquire and process information, however, change significantly across populations. Humans, for instance, demonstrate unique techniques of information transmission \citep{Tomasello2005}. The variety of communication skills and relative brain sizes \citep{Dunbar2007} indicates important constraints on the development of human-like information exchange and processing, which provide superior abilities for joining efforts to control and modify the environment.

Two main factors constrain the evolution of information exchange and intelligent processing. First, information comes at a cost that can be higher than the corresponding evolutionary gain. Many organisms therefore, avoid sustaining excessive sensing and decision making abilities by preferring, for example, stochastic (non-responsive) behavior in a fluctuating environment \citep{Kussell2005,Dubnau2006}. Second, individual fitness can be improved in the course of a competition by selfish hiding of useful information, making cooperative fair signaling evolutionary disadvantageous. Mimicry, deception \citep{Rowell2006} and other methods of selfish information protection are common in the living world, from populations of bacteria \citep{West2006} to states of government. It is however still unclear what allowed humans to overcome them with such an efficiency.

We address here the question: What are the necessary evolutionary selection conditions for the development of communication (significant information exchange and processing) between members of a population? Previous attempts to answer similar questions, can be roughly separated into two main categories: First, analysis of the emergence of various communicative behavior strategies, e.g. cooperation \citep{Fehr2003,Nowak2006,Wilson2007}, fair signaling \citep{Johnstone1998} or acquisition of a common language \citep{Hurford1998,BOYD1985,Cavalli-Sforza1981,Nowak1999a,Nowak1999,Nowak2001}. Second, applying information theory to investigate how signals become associated with meaning \citep{Plotkin2000}. In this work the second approach is followed. The amount of exchanged information is evaluated by the strength of correlations between different signals. This is different from the conventional approach that quantifies only the information content of discrete signals.

Consider interacting organisms generating and displaying some signals, such as information carrying messages or behavior and display patterns. How can one be sure that they communicate, in the sense that some information is transmitted, comprehended and utilized for choosing a specific behavior, e.g. cooperation or selfishness? For instance, no communication is required for interactions where the competitors choose their behavior independently of each other. On the other hand, significant information exchange occurs in the course of correlated reactions between two interacting organisms, that are committed to a common role agreement \citep{Aumann1976}, e.g. who is the donor and who is the recipient. Consequently, correlations between observable behavior patterns can serve as a quantitative measure of information exchange, allowing to construct an ab initio model of communication with no a priori assumptions concerning specific mechanisms of information transfer and processing.

In this paper, we demonstrate that a unique set of evolutionary conditions provides a solution both for a stable, maximal information exchange between individuals and for the naturally occurring diversity of communication abilities across populations. Communication skills in populations are evaluated by presence of correlated, rather than independent, behavior patterns. Correlated behavior is predicted to co-evolve with an ability to affect the competitor's behavior, creating populations with a mixture of evolutionary important correlated interactions of the indirect reciprocity type \citep{Nowak2005}, and more frequent non-responsive behavior, lacking any evolutionary gain. Another result is that fluctuations in population size, which are extreme in human evolution compared to other great apes species \citep{Garrigan2006}, are an indicator of faster evolution of information exchange and processing. The supporting analytical tools are presented in the supplementary material (SM) section.

Any evolvable individual property (a phenotype) emerges as a mutation in a population, subsequently developing or waning with generations as a consequence of natural selection.
The selection process can be approximated by a game \citep{Smith1982,Vincent2005,Nowak2004} where players $A$ and $B$ react mutually. $W_{pq}$ will be associated with the payoff that a player receives for his reaction $p$ vs. $q$ of the competitor. In a given interaction, each competitor may obtain payoffs according to different weight tables. This allows one to take into account that an individual may be found in different modes (states such as e.g. age, intra-population rank, sex, ability to acquire information, etc.) that influence his payoff (see Fig. 1A). Some of the payoff tables have been named by analogy to real-life situations (e.g. Prisoner's Dilemma, Chicken, Battle of the Sexes etc.). It must be stressed though that in these cases both competitors play under identical weight tables (see Fig. 1B).

Consider repeated interactions between two individuals $A$ and $B$. Let us assume that each interaction includes exchange of some information carrying messages of any type, together with either selfish $S$ or cooperation $C$ reactions. The terms "reaction" or "response", rather than strategy, are used to emphasize that individuals do not act independently, but may react to their opponent's actions. The term "strategy" refers here to the decision process that leads to a specific act. For instance, in this context one may say that a player has used the Tit-for-Tat strategy to choose either cooperative or selfish reaction, while playing the Prisoner's Dilemma game.

This paper demonstrates that only the statistics of mutual reactions is required to estimate the amount of information that $A$ and $B$ exchange, comprehend and utilize during their strategic decision making. An alternative approach could be based on analysis of information load of messages generated by the competitors. There exist numerous measures for information in language-like signals, including Shannon entropy \citep{Shannon1948} and Kolmogorov complexity \citep{Kolmogorov1965}. Association of information in a signal with communication, implicitly assumes that competitors fully understand each message. This assumption, however, is not applicable to evolvable systems, since developing signals are always ambiguous.

All information exchanged by competitors $A$ and $B$, pertinent to their mutual behavior, will be associated to the amount of information that one can infer concerning the unknown behavior of $A$ from observing the behavior of $B$ (or vice versa). For instance, an external observer, by observing during an interaction the reaction of $B$, e.g. cooperation, will be able to predict with reduced uncertainty (equals to information gain) the reaction of $A$. Competitors lacking any mutual knowledge, have to exchange information during their interaction, in order to achieve dependent, rather than random, mutual behavior.

Fortunately, information theory provides a standard expression for mutual information \citep{Cover1991}, describing reduction in uncertainty of one signal through the knowledge of another. So far it was applied to analysis of correlations, complexity and communication in physical \citep{Kraskov2004}, linguistic \citep{Pothos2007} and biological systems \citep{Bialek2001,Bednekoff1997,Bulsara1996,Eguia2000}. In our case, mutual information $I$ indicates how much information (in bits) can be derived from the reaction of one of the competitors, while knowing certainly the reaction of the other:
\begin{eqnarray}
I=\sum_{p,q=S,C}\left [\Omega_{pq}\log_{2}\left (\frac{\Omega_{pq}}{{(\Omega_{pS}+\Omega_{pC})(\Omega_{Sq}+\Omega_{Cq})}}\right )\right ],\label{mutinfo1}
\end{eqnarray}
where $\Omega_{pq}$ is the probability of reactions $p$ against $q$. This work derives the selection rules that maximize the value of mutual information $I$ in an evolving system, adopting it as a measure of communication.

The properties of the mutual information expression (\ref{mutinfo1}) justify its relevance to the problem of communication. Mutual information $I$ varies from $0$ to $1$ bit, reaching its maximum in case of positively ($S$ vs. $S$ and $C$ vs. $C$, $\Omega_{SS}=\Omega_{CC}=0.5$ and $\Omega_{SC}=\Omega_{CS}=0$) or negatively ($S$ vs. $C$ and $C$ vs. $S$, $\Omega_{SC}=\Omega_{CS}=0.5$ and $\Omega_{SS}=\Omega_{CC}=0$) correlated reactions, indicating that they require significant information exchange. The minimum values of $I=0$ correspond to independent mutual reactions (as in case of mixed strategies) ($S$ is generated with probability $x_{S}$, $\Omega_{SS}=x^{2}_{S},\Omega_{SC}=x_{S}(1-x_{S}),\Omega_{CS}=(1-x_{S})x_{S},\Omega_{CC}=(1-x_{S})^{2}$), including unconditional selfishness $x_{S}=1$ ($S$ vs. $S$, $\Omega_{SS}=1,\Omega_{SC}=\Omega_{CS}=\Omega_{CC}=0$) and unconditional cooperation $x_{S}=0$ ($C$ vs. $C$, $\Omega_{CC}=1,\Omega_{SC}=\Omega_{CS}=\Omega_{SS}=0$). Independent reactions, indeed, do not require any information exchange. Analysis of the evolutionary stability of populations with correlated reactions is therefore of major importance to describe the evolution of abilities to exchange and process information.

In this model, an individual $A$ possesses several behavior modes (states) $k\in \{1,N\}$, each one characterized by three evolvable parameters $(\epsilon^{A}_{k},\alpha^{A}_{k},\beta^{A}_{k})$ and weight table $W^{k}_{pq}$. In the course of a competition, $A$ acts in a specific mode $k$ with probability $\epsilon^{A}_{k}$, generates either selfish $S$ or cooperative $C$ reactions with a strategy defined by $(\alpha^{A}_{k},\beta^{A}_{k})$ and receives payoffs according to $W^{k}_{pq}$ (Fig. 2A). The conditional probabilities $(\alpha^{A}_{k},\beta^{A}_{k})$ define the statistics of reactions in each behavior mode: $\alpha_{k}$ and $1-\alpha_{k}$ are, respectively, the probabilities to choose cooperation $C$ and selfish $S$ reactions against the competitor's unconditional selfish reaction $S$, while $\beta_{k}$ and $1-\beta_{k}$ are the probabilities to choose cooperation $C$ and selfish $S$ reactions against the competitor's unconditional cooperation reaction $C$ (see Fig. 2B). In this context, $(\alpha^{A}_{k},\beta^{A}_{k})$ can be viewed as selfishness aversion and cooperation attraction correspondingly. Affective behavior is introduced by allowing the phenotypes of competing individuals to depend on each other.

The "mode" index makes possible a multi-game/multi-strategy generalization, taking into account that individuals may possess several behavior states (e.g. age, intra-population rank, sex, ability to acquire information, etc.) that affect both the payoffs $W^{k}_{pq}$ and behavior $(\alpha^{A}_{k},\beta^{A}_{k})$ of an individual $A$ during an interaction. Each mode, therefore, corresponds to a specific strategy and payoff table (see Fig 2A). For instance, when $A$ in $mode(A)$  reacts with $S$ vs. reaction $C$ of $B$ in $mode(B)$,he receives a payoff $W^{mode(A)}_{SC}$ while $B$'s payoff will be $W^{mode(B)}_{CS}$ as shown in Fig 1A. All individuals are assumed to share the same set of behavior modes $k\in {1,N}$.

This phenotype description follows with some alternations previous attempts to incorporate conditional correlations into game theory and biological applications \citep{Hamilton1964,Aumann1974,Eshel1982,Nowak1990,Wahl1999,Skyrms2002,Bergstrom2003,Kussell2005,McNamara2008}.
Conditional probabilities $\alpha$ and $\beta$ are closely related to the reactive strategies approach \citep{Nowak1990,Wahl1999}. Reactive strategies denote the probabilities to cooperate after the competitor has cooperated or defected during the previous interaction, while conditional probabilities assign probabilities for cooperation during the same interaction. This approach makes possible a general description of highly communicative game strategies for choosing a specific reaction (like Tit-for-Tat), together with non-communicative ones (e.g. mixed strategies). Any game strategy in a course of a competition leads to a specific statistics of mutual reactions $\Omega_{pq}$. Conditional probabilities allow describing all possible statistics of reaction, effectively taking into account any game strategy. The individual payoff, defining the evolutionary success, can be expressed though $\alpha$ and $\beta$, since it depends solely on the statistics $\Omega_{pq}$.

The payoff $F(A,B)$ of an individual $A$ interacting with individual $B$ is:
\begin{eqnarray}
F(A,B)=\sum_{i,j}\epsilon^{A}_{i}\epsilon^{B}_{j}\left [\Omega^{i,j}_{SS}W^{i}_{SS}+\Omega^{i,j}_{SC}W^{i}_{SC}+\ldots\right .\nonumber \\
\left .+\Omega^{i,j}_{CS}W^{i}_{CS}+\Omega^{i,j}_{CC}W^{i}_{CC}\right ], \label{payoff1}%
\end{eqnarray}
where the summation includes all possible interactions (mode $i$ of $A$ vs. mode $j$ of $B$), taking into account the corresponding probabilities of such interaction $\epsilon^{A}_{i}\epsilon^{B}_{j}$ and the statistics of the mutual reactions $\Omega^{i,j}_{pq}$. To calculate the payoff $F(A,B)$ and the information exchange rate $I$, one should derive the statistics of the mutual reactions $\Omega^{i,j}_{pq}$ as a function of the competing individuals' phenotypes.

The statistics of the mutual reactions $\Omega^{i,j}_{pq}$ during an interaction of individuals $A$ in mode $i$ vs. $B$ in mode $j$, is a function of the phenotypes $(\alpha_{i},\beta_{i})$, $(\alpha_{j},\beta_{j})$ and their reaction order in the course of a competition (the first to respond lacks confident information on the competitor's intentions). The value of $\Omega^{i,j}_{pq}$ follows from definitions of $\alpha$ and $\beta$, e.g.: $\Omega^{i,j}_{CS}$, the probability of cooperation vs. selfishness in a competition is equal to the conditional probability $\alpha_{i}$ of $A$ to cooperate with a selfish competitor, multiplied by the unconditional probability $\gamma_{ji}$ of $B$ to provide reaction $S$ during a competition with $A$ (see Fig. 2B):
\begin{eqnarray}
\Omega_{SS}=(1-\alpha_{i})\gamma_{ji},\;\Omega_{CC}=\beta_{i}(1-\gamma_{ji}) \nonumber \\
\Omega_{CS}=\alpha_{i}\gamma_{ji},\;\Omega_{SC}=(1-\beta_{i})(1-\gamma_{ji}),
\label{omegaAB}%
\end{eqnarray}
where $\gamma_{ji}$ is:
\begin{eqnarray}
\gamma_{ji}(\alpha_{j},\beta_{j},\alpha_{i},\beta_{i})&=&\nonumber \\
&=&\frac{(1-\beta_{j})-(1-\beta_{i})(\alpha_{j}-\beta_{j})}{{1-(\alpha_{i}-\beta_{i})(\alpha_{j}-\beta_{j})}}. \label{gammaji}%
\end{eqnarray}
Derivation of eq. (\ref{gammaji}) follows from symmetry considerations, for instance the statistics $\Omega^{i,j}_{pq}$ and $\Omega^{j,i}_{pq}$ have to be identical. Fortunately, $\Omega^{i,j}_{pq}$ is independent of the order of reactions in interactions which contribute to the development of communication (see SM \ref{SMsecITIP}, \ref{SMsecECAP} and \ref{SMessb}).

Two individuals possessing the same behavior mode $(\alpha_{i},\beta_{i})=(\alpha_{j},\beta_{j})=(\alpha,\beta)$ exchange during an interaction an amount of information:
\begin{eqnarray}
I(\alpha,\beta)=\log_{2}\left (\frac{(1-\alpha)^{(1-\alpha)\gamma}(1-\beta)^{(1-\beta)(1-\gamma)}\alpha^{\alpha\gamma}\beta^{\beta(1-\gamma)}}{{\gamma^{\gamma}(1-\gamma)^{\gamma}}}\right ),\nonumber \\
\label{mutinfo2}
\end{eqnarray}
where, according to eq. (\ref{gammaji}), $\gamma=\gamma_{ji}(i=j)$ is:
\begin{equation}
\gamma=\frac{1-\beta}{{1+\alpha-\beta}}. \label{gammaii}%
\end{equation}
Eq. (\ref{mutinfo2}) follows from (\ref{mutinfo1}), taking into account eqs. (\ref{omegaAB}) and (\ref{gammaji}) (see SM \ref{SMsubsecIETII}). Expression (\ref{mutinfo2}) provides a quantitative measure of the individual's information processing capability in each behavior mode $(\alpha,\beta)$.

According to eq. (\ref{mutinfo2}), the behavior modes $(\alpha=1,\beta=0)$ and $(\alpha=0,\beta=1)$ possess the maximal information exchange rate ($1$ bit per interaction), as shown in Fig. 2C. These modes can be classified according the statistics of reactions $\Omega_{pq}$ they generate. The mode $(\alpha=0,\beta=1)$ corresponds to positively correlated reactions $(\Omega_{SS}=0.5,\Omega_{CC}=0.5,\Omega_{SC}=\Omega_{CS}=0)$, similar to synchronous interactions of cells in a multicellular organism, e.g. heart contraction. The mode $(\alpha=1,\beta=0)$ corresponds to negatively correlated statistics $(\Omega_{SC}=0.5,\Omega_{CS}=0.5,\Omega_{SS}=\Omega_{CC}=0)$, such as interactions of the indirect reciprocity, taking $C$ as donor and $S$ as recipient). The flux of information is zero for the modes $(\alpha=\beta)$, $(\alpha=0,\beta)$ and $(\alpha,\beta=1)$, corresponding to random mutual reactions, unconditional selfishness $(\Omega_{SS}=1,\Omega_{SC}=\Omega_{CS}=\Omega_{CC}=0)$ and unconditional cooperation $(\Omega_{CC}=1,\Omega_{SC}=\Omega_{CS}=\Omega_{SS}=0)$.

The evolution of positively correlated reactions $(\alpha=0,\beta=1)$ requires much longer times than the evolution of negatively correlated ones $(\alpha=1,\beta=0)$. This is a consequence of the slow evolution near each of the the axes $\alpha=0$ and $\beta=1$, since the interactions in the populations along these axes are all of the unconditional selfishness type $(\Omega_{SS}=1)$ along $\alpha=0$ or cooperation type $(\Omega_{CC}=1)$ along $\beta=1$. There is no evolutionary drive to move between different populations possessing the same statistics of reactions, since they correspond to the same fitness value (\ref{payoff1}). Further analysis, therefore, will focus on the development of communication abilities based on negatively correlated reactions $(\alpha=1,\beta=0)$, leading to interesting analogies with real world phenomena.

A population consisting of phenotypes with a single negatively correlated behavior mode $(\alpha=1,\beta=0)$, is evolutionary unstable under any weight table $W_{pq}$. This can be demonstrated as follows: an individual of this type acts against an identical competitor either as a donor or a recipient with equal probabilities $(\Omega_{SC}=0.5,\Omega_{CS}=0.5,\Omega_{SS}=\Omega_{CC}=0)$, receiving an average payoff of $(W_{SC}+W_{CS})/2$. He demonstrates, however, unconditional cooperation $(\Omega_{CS}\approx 1,\Omega_{SC}\approx\Omega_{SS}\approx\Omega_{CC}\approx0)$ or selfishness $(\Omega_{SC}\approx1,\Omega_{CS}\approx\Omega_{SS}\approx\Omega_{CC}\approx0)$ against mutants described by $(\alpha_{j}=1-\Delta\alpha,\beta_{j}=0)$ and by $(\alpha_{j}=1,\beta_{j}=\Delta\beta)$ $\Delta\alpha,\Delta\beta\ll0$, providing them payoffs $W_{SC}$ and $W_{CS}$. These abrupt changes of behavior are a consequence of the instability of $\gamma_{ji}(\alpha_{j},\beta_{j},\alpha_{i},\beta_{i})$ (\ref{gammaji}) near the point $(\alpha=1,\beta=0)$: $\gamma_{ji}(1,0,1,0)=0.5$, $\lim_{\Delta\alpha\to 0}\gamma_{ji}(1,0,1-\Delta\alpha,0)=0$ and $\lim_{\Delta\beta\to 0}\gamma_{ji}(1,0,1,\Delta\beta)=1$. Either of the mutants $(\alpha_{j}=1-\Delta\alpha,\beta_{j}=0)$ and $(\alpha_{j}=1,\beta_{j}=\Delta\beta)$, therefore, outperforms the $(\alpha=1,\beta=0)$ host, since $W_{SC}>(W_{SC}+W_{CS})/2$ or $W_{CS}>(W_{SC}+W_{CS})/2$. Consequently, evolutionary stability of negative correlations requires phenotypes composed of at least two different behavior modes.

An example of a population with two behavior modes possessing evolutionary stability and diversity of communication skills can be constructed as follows (SM \ref{SMsecSDIR} and \ref{SMsecECAP}). The phenotype of an individual $A$ will include two behavior modes (responsive $(R)$ and non-responsive $(NR)$), described by three parameters $(\alpha_{A},\beta_{A},\epsilon_{A})$, where $\epsilon_{A}$ takes now the meaning of affect ability. In the course of an interaction, the individuals $A$ and $B$ possess non-responsive behavior mode $(NR)$ with probabilities defined by the affect abilities of the competitor, meaning that non-responsive behavior is not a personal choice, but enforced by the opponent.
In case of $A$, therefore, $1-\epsilon_{B}$ and $\epsilon_{B}$ are the probabilities to act in responsive $(R)$ and non-responsive $(NR)$ modes correspondingly (Fig. 3). In the non-responsive mode, an individual behaves stochastically, generating reactions with probabilities matching the average statistics of its reactions, and receives no payoff:
\begin{eqnarray}
W^{NR}_{pq}=0. \label{Stabcondzero}
\end{eqnarray}
In the responsive mode, its behavior is defined by the conditional probabilities $(\alpha,\beta)$, and, therefore, depends on the reactions of the competitor. The lower the number of pairwise interactions in this mode, the greater the weight for each competition:
\begin{eqnarray}
\widetilde{W}^{R}_{pq}\propto\frac{1}{{1-\epsilon_{B}}}\times\begin{tabular}{c|c|c}
  & $C$ & $S$ \\
\hline
$C$ & $W_{CC}$ & $W_{CS}$ \\
\hline
$S$ & $W_{SC}$ & $W_{SS}$  \\
\end{tabular}. \label{Stabcondnonrandom}
\end{eqnarray}
The interactions of a population with high affect abilities $\epsilon\rightarrow 1$ are composed, therefore, of a small amount $(\propto 1-\epsilon)$ of evolutionary important competitions with negatively correlated reactions and significant information exchange, together with more frequent non-responsive behavior lacking any evolutionary gain (Fig. 3).

In such a population, communication abilities will evolve when $W_{pq}$ (see \ref{Stabcondnonrandom}) fluctuate around the values corresponding to games of Chicken, Leader and Battle of the Sexes (see Fig. 1B and SM \ref{SMsecGFDD}). No development occurs either when $W_{pq}$ do not fluctuate or correspond to other games, e.g. Prisoner's Dilemma. To describe the evolutionary dynamics, the instantaneous state of a population is defined as a density distribution over the phenotype space $\rho(\overrightarrow{r})=(\alpha,\beta,\epsilon)$. The time development is governed by the replicator dynamics equations \citep{Taylor1978}: it consists of the emergence of mutations due to a diffusion-like process and redistribution between the old phenotypes according to natural selection:
\begin{equation}
\frac{\partial\rho(\overrightarrow{r})}{{\partial t}}=\rho(\overrightarrow{r})\frac{F(\overrightarrow{r})-\overline F}{{T_{gen}|\overline F|}}+D\frac{\partial^{2}\rho}{{\partial \overrightarrow{r}^{2}}},\;\bar F= \int\rho(\overrightarrow{r})F(\overrightarrow{r})d\overrightarrow{r}, \label{repdyn}%
\end{equation}
where $F$ denotes the fitness, $\bar F$ the average fitness, $D$ the diffusion coefficient of the mutation process and $T_{gen}$ the time-span of a single generation. These equations can be reduced to the local velocities of a population in the phenotype space (SM \ref{EPP}). In the case of a particle-like population confined to the vicinity of its average phenotype one obtains, taking condition (\ref{Stabcondzero}) into account:
\begin{equation}
v_{\alpha}=\frac{\overline{\left | \Delta\alpha \right |^{2}}}{{T_{gen}F(\overrightarrow{r})}}\frac{\partial F(\overrightarrow{r})}{{\partial\alpha}},\; v_{\beta}=\frac{\overline{\left | \Delta\beta \right |^{2}}}{{T_{gen}F(\overrightarrow{r})}}\frac{\partial F(\overrightarrow{r})}{{\partial\beta}}, \label{valalpbeta}
\end{equation}
\begin{equation}
v_{\epsilon}\propto\frac{\overline{\left | \Delta\epsilon \right |^{2}}}{{T_{gen}F(\overrightarrow{r})}}\left [\frac{\partial F(\overrightarrow{r})}{{\partial\alpha}}\frac{\partial^{2} F(\overrightarrow{r})}{{\partial\epsilon\partial\alpha}}+\frac{\partial F(\overrightarrow{r})}{{\partial\beta}}\frac{\partial^{2} F(\overrightarrow{r})}{{\partial\epsilon\partial\beta}}\right ], \label{valeps}
\end{equation}
where $\overline{\left | \Delta\alpha \right |^{2}}$, $\overline{\left | \Delta\beta \right |^{2}}$ and $\overline{\left | \Delta\epsilon \right |^{2}}$ are the spreads of the population along $\alpha$, $\beta$ and $\epsilon$ directions correspondingly. The development of affect abilities $\epsilon$ depends on the changes in $\alpha$ and $\beta$ $(v_{\epsilon}\propto \left |(v_{\alpha},v_{\beta})\right |)$. Populations, therefore, can possess many different values of affect abilities ($v_{\epsilon}=0$ for any value of $\epsilon$) and corresponding information skills at the stable points $v_{\alpha}=v_{\beta}=0$, matching the diversity of communication abilities in nature (see Figs. 4A-4D).

The stronger the fluctuations of the size of a population, the faster the development time of communication abilities in a high affect, negatively correlated behavior mode: $(\epsilon\rightarrow 1,\alpha=1,\beta=0)$ (Fig. 4E):
\begin{equation}
T_{dev}\propto T_{gen}\left (DT_{gen} \right)^{-\frac{3}{{4}}}\overline{\left (\frac{\Delta S}{{S}}\right )^{2}}^{-\frac{1}{{2}}}, \label{timedev}%
\end{equation}
since fluctuations of the payoffs $W_{pq}$ correspond to fluctuations in the size $S$ of a population. The time dependent evolutionary conditions keep the population away from its stable points $(v_{\alpha}=v_{\beta}=v_{\epsilon}=0)$, leading to an increase of the affect abilities $(v_{\epsilon}>0)$ and evolutionary stability of the maximum possible information exchange in the indirect reciprocity behavior mode $(\epsilon\rightarrow 1,\alpha=1,\beta=0)$. Chicken and Battle of the Sexes games possess the fastest development rate of communication abilities (see Fig. 4F).

The prediction that development of communication is accelerated with population size variations, is supported by evidence of significant demographic fluctuations during human evolution \citep{Garrigan2006}. In addition, the long time required to develop multicellularity on Earth (about $75\%$ of the entire evolution time-span) \citep{Hedges2004,Stearns2007} can be associated with the slow development of positively correlated synchronous behavior mode $(\alpha=0,\beta=1)$. Negatively correlated reactions of the indirect reciprocity type $(\alpha=1,\beta=0)$ lack selfish competitions $(\Omega_{SS}=0)$ and match, therefore, the observations of relatively low aggression levels in humans \citep{Wrangham2006}. From a single interaction's perspective, indirect reciprocity can be interpreted as altruistic. It allows one to suggest that non-kin altruism, considered to be a unique human property \citep{Gintis2003}, could be a by-product of evolution of communication abilities.

The derived conditions are essential for evolution and diversity of communication (SM \ref{SMsecSDIR}). Deviations from the scaling of the payoffs per behavior mode (\ref{Stabcondnonrandom}) (e.g. scaling per pairwise interactions) destroy the evolutionary stability of the high affect, negatively correlated $(\epsilon\rightarrow 1,\alpha=1,\beta=0)$ behavior mode. Non-responsive behavior with finite, rather than zero (\ref{Stabcondzero}), payoffs prevents multi-stability (diversity) of the communication development. The inability to affect a competitor to behave in a specific behavior mode makes evolution of the information exchange and processing impossible. The same statements are valid in the case of multi-behavior mode phenotypes. Non-diverse (all populations converge to the same state) development of negative correlations, however, is possible, raising the question whether there exists a scheme for evolution of information exchange, which is different from the emergence of human-like communication abilities?

The results of this work indicate that development of artificial life or intelligence may be impossible without selection rules changing with time. The attempts to create artificial evolution of competing digital organisms led so far to a limited, rather than unbounded, growth of complexity \citep{Ray1994,Wilke2001}. Evolution of communication can be considered as a base for significant complexity growth in living and artificial systems. The analogies of these results with human evolution (e.g. presence of demographic bottlenecks) indicate a promising direction for the artificial evolution of intelligent systems composed of communicating agents.

The developed framework does not predict exact mechanism for the evolution of communication. Only general evolutionary conditions are derived that lead to correlated statistics of reactions, assuming that the mechanisms for the necessary information exchange and processing can be implemented in various ways and are species' dependent.

To conclude, a mathematical framework based on observable, correlated behavior patterns was developed for analysis of evolution of information exchange in populations. It predicts that the communication abilities of populations are related to negatively correlated interactions (such as donor vs. recipient) and affective behavior. It was found that fluctuations in population size accelerate significantly the evolution of these abilities, allowing thus to explain the uniqueness of the human-line evolution. Competitions similar to games of Chicken and Battle of the Sexes lead to a faster evolution of communication. An experimental verification of models based on this framework is in principle possible, since the developed analytical tools are based on observable parameters, such as statistics of mutual reactions during competitions, rather than exact modeling of information exchange and decision making biological mechanisms.


\section*{Acknowledgments}

I owe a great debt to A. Englander for his help in all stages of this work. Helpful discussions with  O.
Khasanov, A. Morozov and M. Narovlyansky are gratefully
acknowledged.



\begin{widetext}
\clearpage
\begin{figure}
  \begin{center}
    \begin{tabular}{c c}
      \multicolumn{1}{l}{{\bf\sf A}} & \multicolumn{1}{l}{{\bf\sf B}}\\
      \resizebox{0.333\textwidth}{!}{\includegraphics{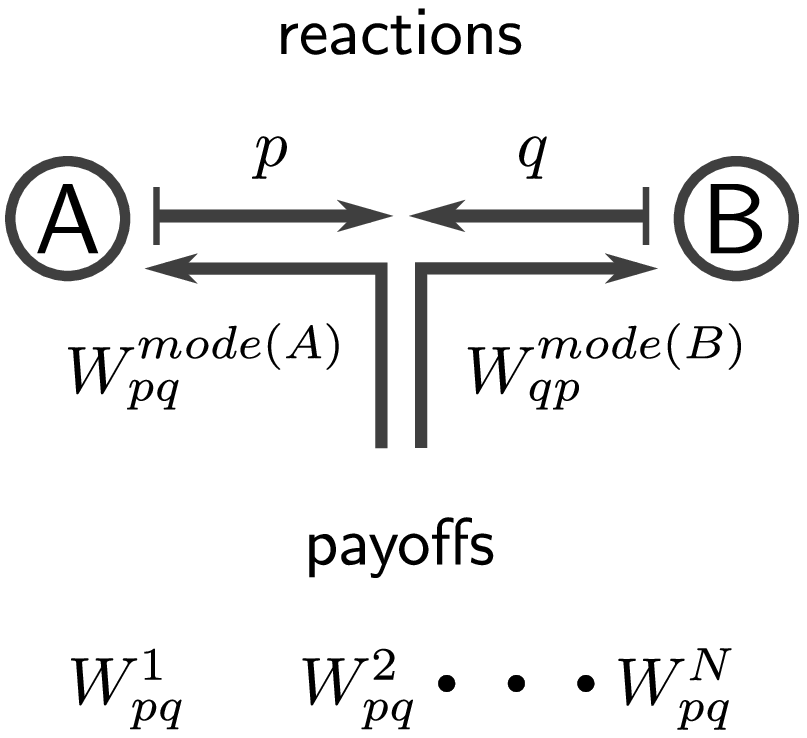}} &
      \resizebox{0.333\textwidth}{!}{\includegraphics{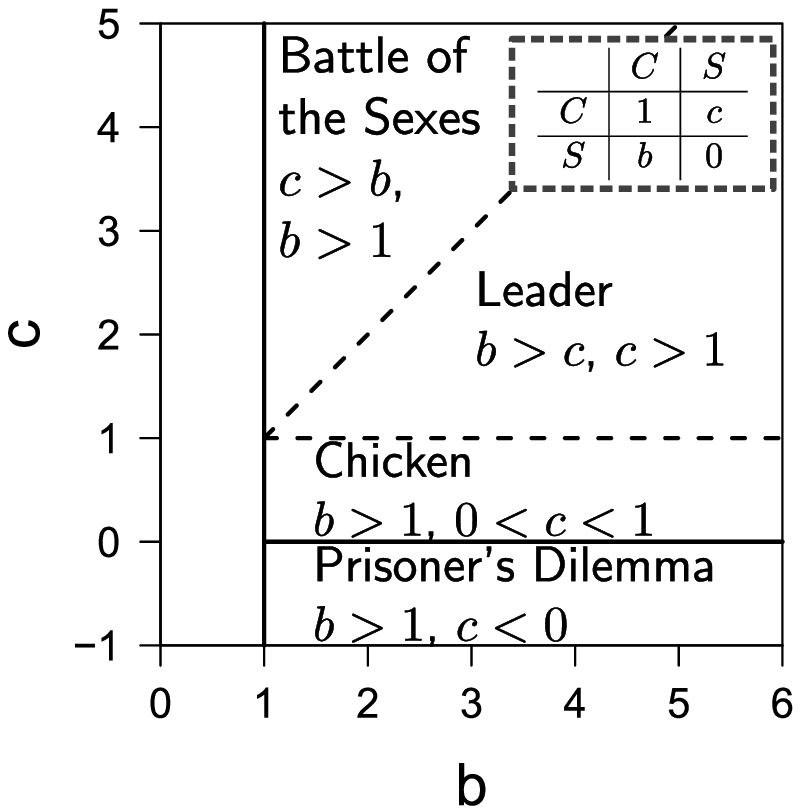}} \\
    \end{tabular}
    \caption{Evolution as a game. ({\bf A}) Interaction of individuals $A$ and $B$, possessing behavior modes $mode(A)$ and $mode(B)$ correspondingly. $A$ and $B$ generate reactions $p$ and $q$, receiving payoffs according to tables specific to each mode, e.g. $W^{mode(A)}_{pq}$ is the payoff of $A$ for its reaction $p$ vs. reaction $q$ of its competitor. Indices $p$ and $q$ represent either selfish $S$ or cooperative $C$ reactions. Different modes correspond to the possible states of the competitors (e.g. age, intra-population rank, sex, ability to acquire information and etc. \ldots) that determine the individual payoffs $W^{k}_{pq}$ during an interaction. All individuals are assumed to share the same set of modes $k\in \{1,N\}$. ({\bf B}) Classification of payoff values $(b,c)$ according to evolutionary game theoretical models, assuming that the competing individuals are in the same mode, with payoffs $W_{SS}=0$, $W_{SC}=b$, $W_{CS}=c$ and $W_{CC}=1$. Payoff values within the unmarked areas lack specific names.}
    \label{fig1}
  \end{center}
\end{figure}
\clearpage
\begin{figure}
  \begin{center}
    \begin{tabular}{c c c}
      \multicolumn{1}{l}{{\bf\sf A}} & \multicolumn{1}{l}{{\bf\sf B}} & \multicolumn{1}{l}{{\bf\sf C}}\\
      \resizebox{0.333\textwidth}{!}{\includegraphics{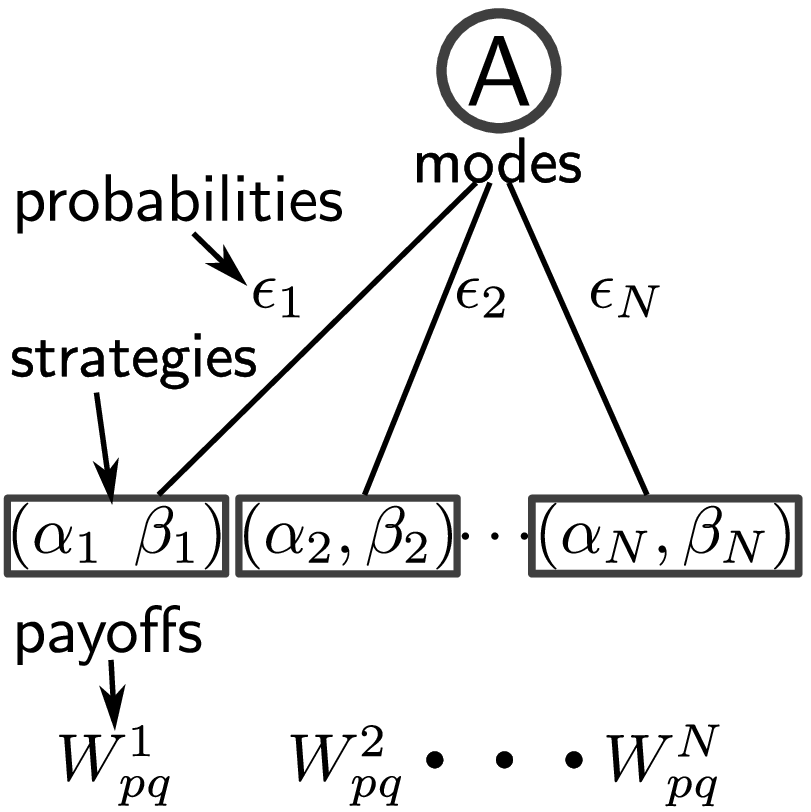}} &
      \resizebox{0.333\textwidth}{!}{\includegraphics{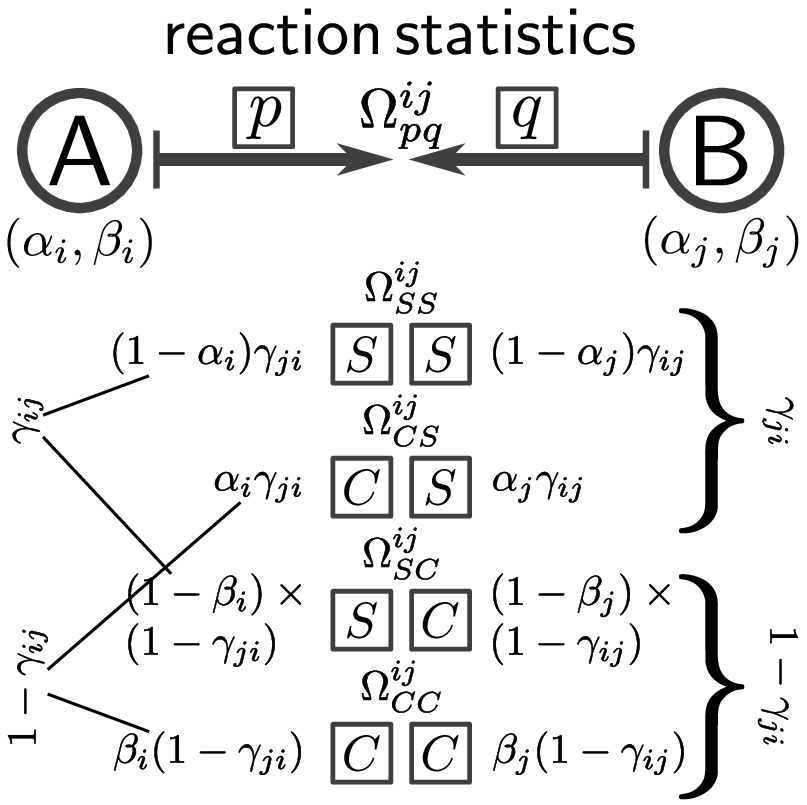}} &
      \resizebox{0.333\textwidth}{!}{\includegraphics{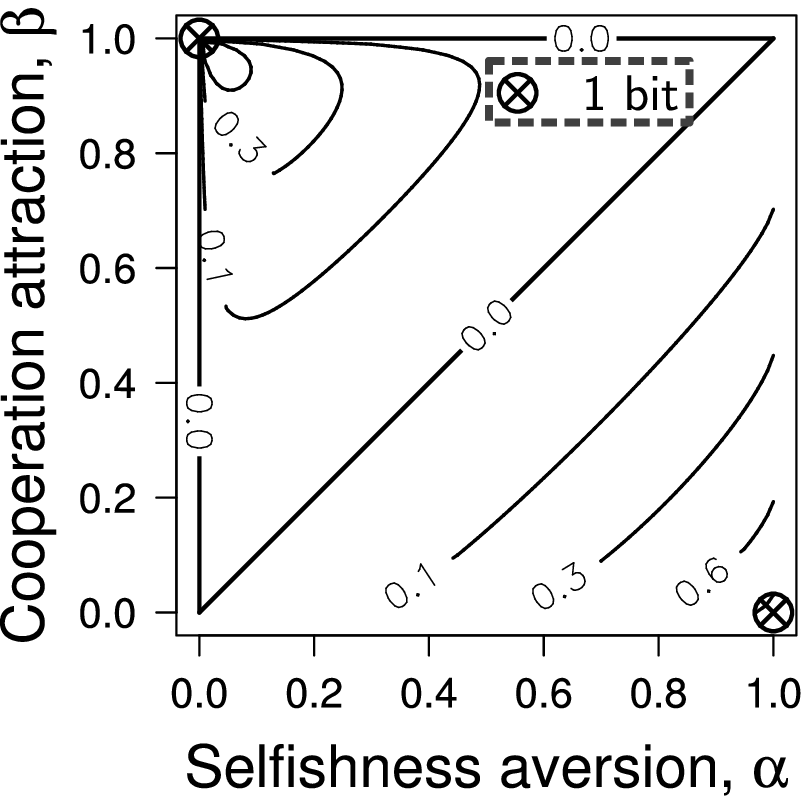}} \\
    \end{tabular}
    \caption{Correlations of mutual reactions and information exchange in the course of a competition. ({\bf A}) The strategy of an individual $A$ in behavior mode $i$ is defined by $\alpha_{i}$ and $\beta_{i}$: $\alpha$ denotes the conditional probability of player's A cooperation reaction $C$ vs. the selfish reaction $S$ of its competitor $B$, whereas $\beta$ denotes the conditional probability for cooperation $C$ vs. cooperation $C$ of the competitor. Selfishness aversion and cooperation attraction are possible interpretations for $\alpha$ and $\beta$. ({\bf B}) The statistics $\Omega^{ij}_{pq}$ for reaction $p$ by $A^{mode(i)}$ vs. $q$ by $B^{mode(j)}$ follows from the definitions of $\alpha$ and $\beta$ (eq. (\ref{omegaAB})). To determine the probability $\Omega^{ij}_{CS}$ of $C$ vs. $S$ interaction, for instance, one must multiply the conditional probability $\alpha_{i}$ of $A$ to cooperate with a selfish competitor $B$ by the unconditional probability $\gamma_{ji}$ of $B$ to provide reaction $S$ during a competition with $A$. The values of unconditional probabilities $\gamma_{ji}$ and $\gamma_{ij}$ follow from symmetry considerations (eq. (\ref{gammaji})). ({\bf C}) Contour plot of information exchange per interaction $I(\alpha,\beta)$ in populations consisting of individuals competing in the same mode $(\alpha,\beta)$. The diagonal $(\alpha=\beta)$ together with the axes $(\alpha=0)$ and $(\beta=1)$ correspond to a minimum of $0$ bits exchanged per interaction matching the cases of random mutual reactions, of unconditional cooperation $(\beta=1 | \Omega_{CC}=1,\Omega_{SC}=\Omega_{CS}=\Omega_{SS}=0)$ and of unconditional selfishness $(\alpha=0 | \Omega_{SS}=1,\Omega_{SC}=\Omega_{CS}=\Omega_{CC}=0)$. A maximum of $1$ bit per interaction is achieved at the points $(\alpha=0,\beta=1)$ and $(\alpha=1,\beta=0)$ (marked with a $\bigotimes$), corresponding to the synchronous $(\Omega_{SS}=0.5,\Omega_{CC}=0.5,\Omega_{SC}=\Omega_{CS}=0)$ and indirect reciprocity $(\Omega_{SC}=0.5,\Omega_{CS}=0.5,\Omega_{SS}=\Omega_{CC}=0)$ behavior modes. Tiny deviations from the synchronous point $(\alpha=0,\beta=1)$ will bring the mutual information exchange to $0$, due to the very steep gradient of $I(\alpha,\beta)$ in this region.}
    \label{fig2}
  \end{center}
\end{figure}
\clearpage
\begin{figure}
  \begin{center}
      \resizebox{0.333\textwidth}{!}{\includegraphics{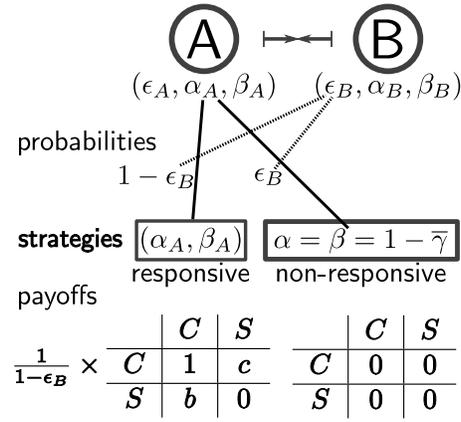}} 
    \caption{Essential conditions for evolution and diversity of communication in a population with two behavior modes per individual. An individual $A$ is either in a non-responsive or a responsive mode, with the probabilities $\epsilon_{B}$ and $1-\epsilon_{B}$, defined by the affect ability $\epsilon_{B}$ of its competitor $B$. The affect ability $\epsilon$ may be considered as an individual's skill to influence the behavior of its competitor, enforcing him in this way to behave stochastically. There is zero payoff for non-responsive behavior $(W_{pq}=0)$. The lower the probability to be in responsive mode, the greater the payoff during a competition. In a population with developed affect abilities $(\epsilon \rightarrow 1)$, the responsive behavior corresponds to high evolutionary payoffs, but interactions of this type are rare. In the non-responsive mode an individual generates arbitrary reaction with probabilities matching the average statistics of its reactions. Non-responsive behavior can be interpreted as a reaction induced by a signal from an arbitrary member of the population, rather than from a competitor. The reaction in this case matches the statistics of the mean reactions, since the signal effectively averages over all player's possible interactions within the population.}
    \label{fig3}
  \end{center}
\end{figure}
\clearpage
\begin{figure}
  \begin{center}
    \begin{tabular}{c c c}
      \multicolumn{1}{l}{{\bf\sf A}} & \multicolumn{1}{l}{{\bf\sf B}} & \multicolumn{1}{l}{{\bf\sf C}} \\
      \resizebox{0.3\textwidth}{!}{\includegraphics{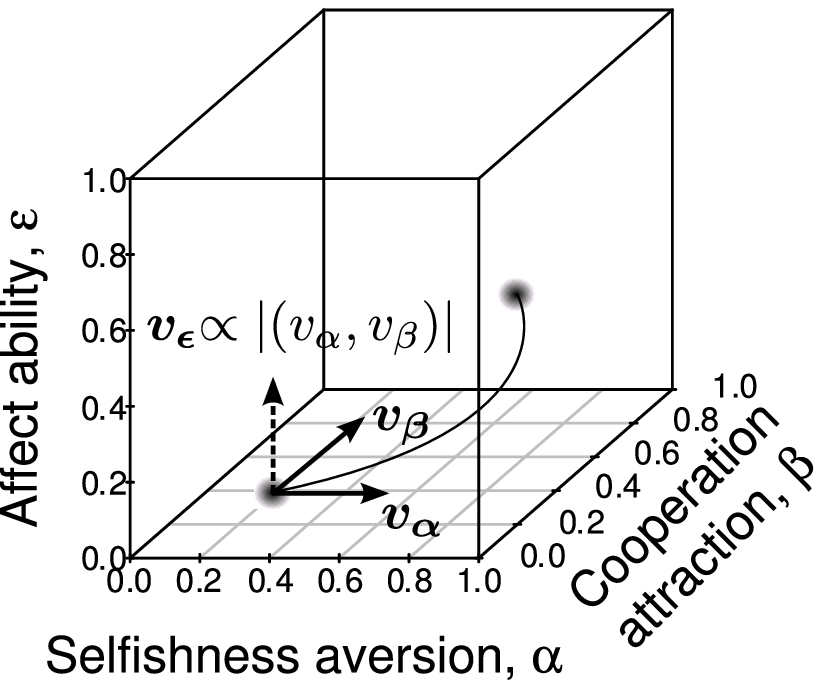}} &
      \resizebox{0.3\textwidth}{!}{\includegraphics{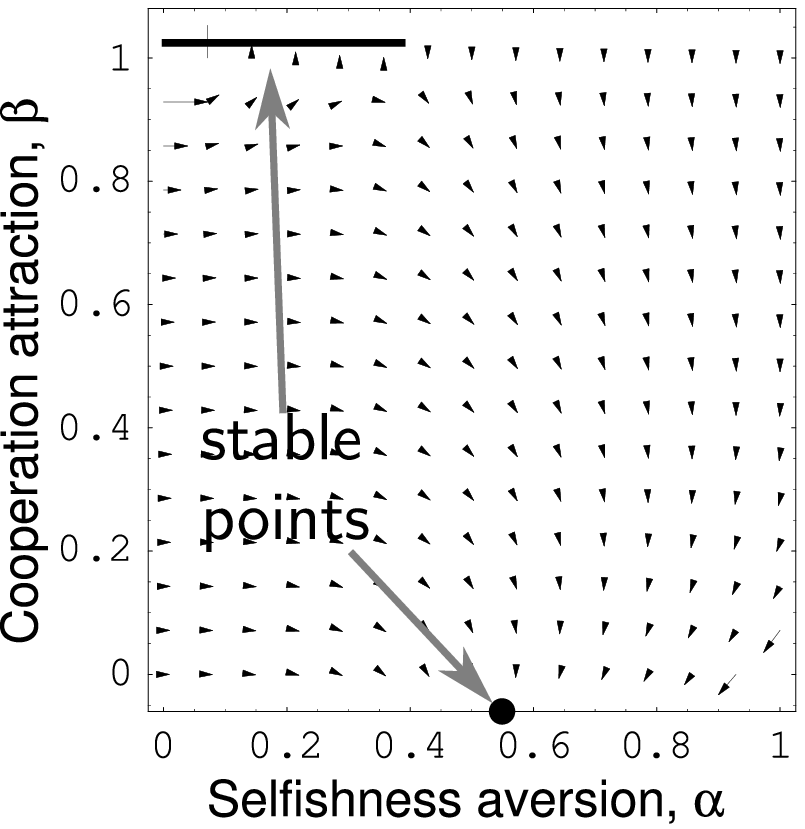}} &
      \resizebox{0.3\textwidth}{!}{\includegraphics{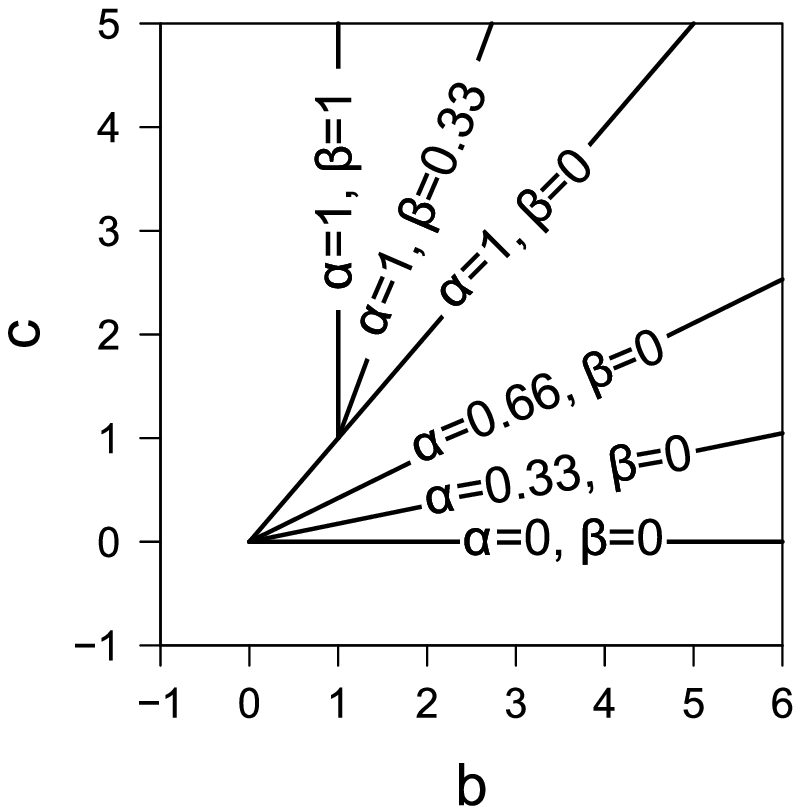}} \\
      \multicolumn{1}{l}{{\bf\sf D}} & \multicolumn{1}{l}{{\bf\sf E}} & \multicolumn{1}{l}{{\bf\sf F}} \\
      \resizebox{0.3\textwidth}{!}{\includegraphics{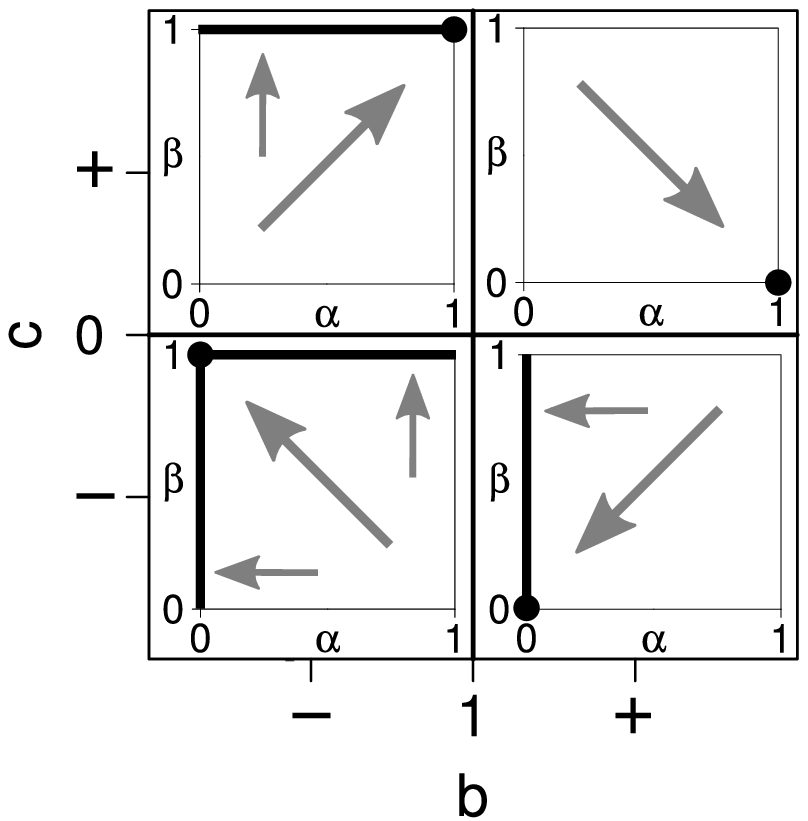}} &
      \resizebox{0.3\textwidth}{!}{\includegraphics{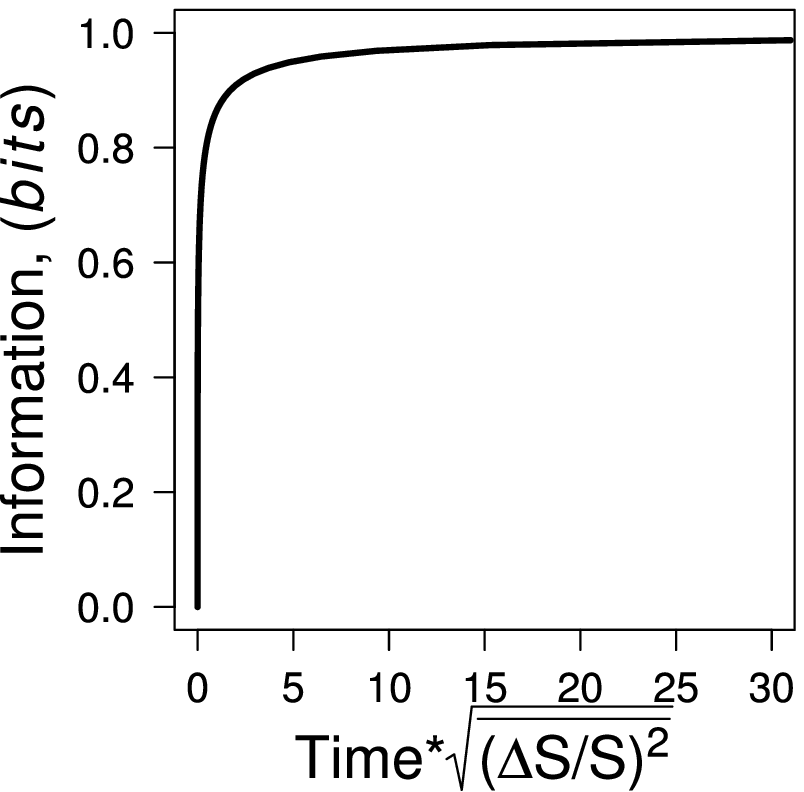}} &
      \resizebox{0.3\textwidth}{!}{\includegraphics{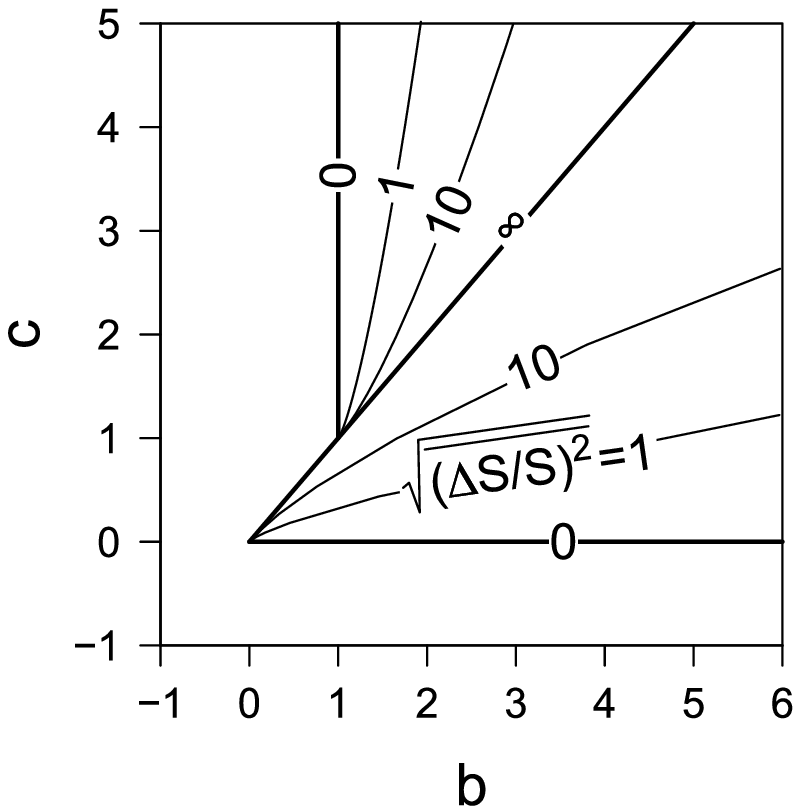}} \\
    \end{tabular}
    \caption{Development of communication abilities as a consequence of fluctuations in population size. ({\bf A}) For the population presented in Fig. 3., the local velocity $\vec{v}(\epsilon,\alpha,\beta,b,c) = (v_{\alpha},v_{\beta},v_{\epsilon})$ defines the instantaneous evolution of a confined population in the phenotype space $(\alpha,\beta,\epsilon)$. Development of affect abilities $v_{\epsilon}=\partial \epsilon/t$ is proportional to $v_{\alpha}$ and $v_{\beta}$, the changes with time of selfishness aversion $\alpha$ and cooperation attraction $\beta$ correspondingly. ({\bf B}) The vector field of $\vec{v}$ in case of $(\epsilon=0.5,b=1.5,c=0.5)$. A population converges to one of its stable states with $v_{\alpha}=v_{\beta}=v_{\epsilon}=0$. Lines and points of convergence are marked in bold. ({\bf C}) The stable points $(\alpha_{st},\beta_{st})$ in the case $\epsilon=0.5$, as a function of the payoffs $(b,c)$. ({\bf D}) Convergence of a population towards its stable points for $\epsilon\rightarrow 1$, as a function of payoffs $(b,c)$. Significant information exchange at the indirect reciprocity behavior mode $(\alpha=1,\beta=0)$ is stable for Chicken, Leader and Battle of the Sexes games (see Fig. 1B). ({\bf E}) Development of communication with time in case of fluctuating payoffs $(\overline{b}=1.5,\overline{c}=0.5,\sqrt{\overline{\Delta b^{2}}}=\sqrt{\overline{\Delta c^{2}}}\approx\sqrt{\overline{(\Delta S/S)^{2}}}$), where $S$ is size of the population. Fluctuation of the payoffs $(b,c)$, associated with the fast fluctuations in size of the population $\sqrt{\overline{(\Delta S/S)^{2}}}$, accelerate the development of maximal information exchange at the indirect reciprocity behavior mode $(\alpha=1,\beta=0)$, changing the corresponding development timescale. ({\bf F}) Population size fluctuations $\sqrt{\overline{(\Delta S/S)^{2}}}$ leading to equal development rate of $I$ $(\partial I/\partial t\propto v_{\epsilon})$ at $\epsilon=0.5$ for different values of payoffs $(b,c)$. For small population size fluctuations the fastest information exchange rate will occur with Chicken and Battle of the Sexes games.}
    \label{fig4}
  \end{center}
\end{figure}
\clearpage
\end{widetext}

\clearpage
\appendix

\section{Supplementary materials for A. Feigel "Conditions required for the evolution of communication abilities within a species"}
This supplement includes a detailed description of the analytical results presented in the main text. Analysis of evolutionary stability of populations with significant information exchange between its members comes after an introduction to the evolutionary game theory. The evolutionary conditions required for the development and diversity of information exchange, as well as the corresponding population dynamics, are derived.
\subsection{Evolutionary game theory}
\subsubsection{Selection rules as a game}
In the framework of evolutionary game theory, a competition of two individuals is presented as a game. For two-reaction games, the selection rules can be presented by payoff tables:
\begin{equation}
\begin{tabular}{c|c|c}
  & $C$ & $S$ \\
\hline
$C$ & $W^{mode}_{CC}$ & $W^{mode}_{CS}$ \\
\hline
$S$ & $W^{mode}_{SC}$ & $W^{mode}_{SS}$  \\
\end{tabular} \label{SMWtable1}
\end{equation}
where $W^{mode}_{pq}$ is the payoff for reaction $p$ made in the specific behavior $mode$ against a competitor's reaction $q$. We assume that the payoffs of an individual during an interaction depend on its $mode$, corresponding to different individual states (e.g. age, intra-population rank, sex, ability to acquire information and etc. \ldots) affecting the personal payoffs during an interaction.

Reactions $S$ and $C$ can be defined as selfish and cooperative correspondingly, assuming $W^{mode}_{SS}<W^{mode}_{CC}$.

The total gain $G(m,h)$ of an individual $m$ for an interaction with an individual $h$ is:
\begin{equation}
G(m,h)=\sum_{i,j}\Xi(m_{i},h_{j})P(m_{i}|h_{j}|W^{i}_{pq}), \label{SMpayoff0}%
\end{equation}
where $\Xi(m_{i},h_{i})$ is the probability for the individual $m$ to be in the mode $i$ while its competitor $h$ is in the mode $j$ and $P(m_{i}|h_{j}|W^{i}_{pq})$ is the payoff of $m$:
\begin{eqnarray}
P(m_{i}|h_{j}|W^{i}_{pq}) &=& \Omega_{SS}(m_{i},h_{j})W^{i}_{SS}+\Omega_{SC}(m_{i},h_{j})W^{i}_{SC}+\nonumber \\
&&+\Omega_{CS}(m_{i},h_{j})W^{i}_{CS}+\Omega_{CC}(m_{i},h_{j})W^{i}_{CC},\nonumber \\ \label{SMpayoff1}%
\end{eqnarray}
where $\Omega_{pq}(m_{i},h_{j})$ is the probability for individual $m$ in mode $i$ to provide reaction $p$ against reaction $q$ of its competitor $h$ in mode $j$.

The fitness $F(m)$ of an individual $m$ is given by:
\begin{equation}
F(m)=\sum_{h} \rho(h)G(m,h), \label{SMfitness}%
\end{equation}
where $\rho(h)$ is the density of individuals $h$ in the population and summation goes over all individuals with distinct properties (phenotypes).

\subsubsection{Dynamics and evolutionary stability of a population}
A population can be defined as a density distribution $\rho(\xi)$ over all possible phenotypes $\xi$. The density $\rho(\xi)$, therefore, corresponds to the relative amount (frequency) of phenotypes $\xi$ in the population. Evolution of a population consists of two processes: redistribution between existing phenotypes from generation to generation (natural selection) and emergence of new phenotypes (mutations).

The natural selection is described by the replicator dynamics equations \citep{Taylor1978}, assuming that the fitness is a measure of the progeny:
\begin{equation}
\frac{\partial\rho(\xi)}{{\partial t}}=\frac{1}{{T_{gen}}}\rho(\xi)\frac{F(\xi)-\overline F}{{|\overline F|}}+\widetilde{M}, \label{SMrepdyn}%
\end{equation}
where $T_{gen}$ is the time span of a single generation, $F(\xi)$ is the evolutionary fitness (\ref{SMfitness}) of the phenotype $\xi$ and $\widetilde{M}$ is a contribution of mutations to the density $\rho(\xi)$ of the phenotype $\xi$. The average fitness $\overline F$ in the population is:
\begin{equation}
\overline F= \sum_{\xi=1}^N \rho(\xi)F(\xi). \label{SMfitnessaverage}%
\end{equation}
The difference between the fitness $F(\xi)$ and the average fitness $\overline F$ defines the growth ($F(\xi)>\overline F$) or decay ($F(\xi)<\overline F$) of the corresponding phenotype density $\rho(\xi)$.

The evolutionary stable phenotype $\xi(st)$ has maximum fitness, meaning:
\begin{equation}
F(\xi(st))>F(\chi), \label{SMstabcond0}%
\end{equation}
for all $\chi\neq \xi(st)$. In the case of a population composed of two phenotypes (mutant and evolutionary stable host), the condition (\ref{SMstabcond0}) reduces to:
\begin{equation}
G(host,host)>G(mut,host), \label{SMstabcond1}%
\end{equation}
where $G(m,h)$ is the average payoff for an interaction between phenotypes $m$ and $h$ (see eq. (\ref{SMpayoff0})).

To obtain condition (\ref{SMstabcond1}), let us consider a mutant with a density ($\rho(mut)=\Delta\rho$) invading a population composed of some host phenotype ($\rho(host)=1-\Delta\rho$). Taking eq. (\ref{SMfitness}) into account, the finesses of the host and the mutant are:
\begin{equation}
F(host)=(1-\Delta\rho)G(host,host)+\Delta\rho G(host,mut), \label{SMhostfit}%
\end{equation}
\begin{equation}
F(mut)=(1-\Delta\rho)G(mut,host)+\Delta\rho G(mut,mut). \label{SMmutantfit}%
\end{equation}
In the limit of $\Delta\rho << 1$, conditions (\ref{SMstabcond0}) ($F(host)>F(mut)$) and (\ref{SMstabcond1}) converge to each other.

In the case of equality $G(host,host)=G(mut,host)$ the mutant can become evolutionary beneficial in case $G(host,mut)<G(mut,mut)$. The population dynamics in this case is slow, depending on rare ($\propto \Delta\rho^{2}$) mutant vs. mutant interactions.

\subsection{Evolution of a population as a particle-like motion}
\label{EPP}
In this work, whenever possible, the evolutionary dynamics is going to be considered in the limit of a confined population, composed of a single phenotype with small variations. We are going to develop the particle-like equations of motion of such population in the phenotype space, under assumption that the mutations (see (\ref{SMrepdyn})) correspond to a diffusion-like process:
\begin{equation}
\widetilde{M}=D\frac{\partial^{2}\rho}{{\partial \overrightarrow{r}^{2}}}, \label{SMmutdiff}%
\end{equation}
where $D$ is a diffusion coefficient and $\overrightarrow{r}$ is the phenotype space coordinate. For small $D$, mutations cause only minute changes in the existing phenotypes and a population remains, therefore, confined during its entire evolution. The motion of a confined population can be described by local velocities (a flow field) in the phenotype space, similar to the motion of a particle.

The dynamics of a population (eq. (\ref{SMrepdyn})), in case of a diffusion-like mutation process (eq. (\ref{SMmutdiff})), is described by:
\begin{equation}
\frac{\partial\rho(\overrightarrow{r})}{{\partial t}}=\frac{1}{{T_{gen}}}\rho(\overrightarrow{r})\frac{F(\overrightarrow{r})-\overline F}{{|\overline F|}}+D\frac{\partial^{2}\rho}{{\partial \overrightarrow{r}^{2}}}, \label{SMrepdyn1}%
\end{equation}
\begin{equation}
F(\overrightarrow{r})=\int d\overrightarrow{r}'\rho(\overrightarrow{r}')G(\overrightarrow{r},\overrightarrow{r}'), \label{SMfitcont}%
\end{equation}
\begin{equation}
\overline F=\int d\overrightarrow{r}d\overrightarrow{r}'\rho(\overrightarrow{r})\rho(\overrightarrow{r}')G(\overrightarrow{r},\overrightarrow{r}'), \label{SMavfitcont}%
\end{equation}
where an instantaneous state of the evolving population is defined by a density distribution $\rho(\overrightarrow{r})$ over some $N$-dimension phenotype space $\overrightarrow{r}=(x_{1},x_{2},...,x_{N})$. The fitness $F(\overrightarrow{r})$ and the average fitness $\overline F$ are described by the continuous versions of eqs. (\ref{SMfitness}) and (\ref{SMfitnessaverage}).

A confined population is described by its spread (size in the phenotype space) and by its asymmetry along the direction of the propagation. Its dynamics, therefore, can be approximated by a $\delta$-function, taking into account both the spread $\delta^{(2)}$ (the second derivative) and the asymmetry $\delta^{(1)}$(the first derivative):
\begin{eqnarray}
\rho(\overrightarrow{r},t) &=& \delta(\overrightarrow{r}-\overrightarrow{r}_{0}(t))-\nonumber \\
&&-\overline{x}(t)\delta^{(1)}(\overrightarrow{r}-
\overrightarrow{r}_{0}(t))+
\frac{1}{{2}}\overline{x^{2}}(t)\delta^{(2)}(\overrightarrow{r}-\overrightarrow{r}_{0}(t)),\nonumber \\
\label{SMrhodelta}
\end{eqnarray}
where $\overrightarrow{r}_{0}(t)$ is the location of the population, $\overline{x_{i}}=\int \rho (x_{i}-x_{0i})$ corresponds to the asymmetry and $\overline{x^{2}_{i}}=\int \rho (x_{i}-x_{0i})^{2}$ contributes to the spread $(\overline{x^{2}_{i}}-\overline{x_{i}}^{2})$.

The velocity of a population $\frac{\partial \overrightarrow{r}_{0}}{{\partial t}}$ is:
\begin{equation}
\overrightarrow{v}=\overrightarrow{v^{1st}}+\overrightarrow{v^{2nd}}, \label{SMvival}%
\end{equation}
where the first $\overrightarrow{v^{1st}}$ and the second $\overrightarrow{v^{1st}}$ orders are:
\begin{widetext}
\begin{eqnarray}
v^{1st}_{i}(\overrightarrow{r}) &=& \frac{1}{{T_{gen}}}\frac{1}{{G(\overrightarrow{r},\overrightarrow{r})}}\frac{\partial G(\overrightarrow{r},\overrightarrow{r}')}{{\partial x_{i}}}\left ( \overline{x^{2}_{i}}-\overline{x_{i}}^{2} \right ),\label{SMvival1}
\end{eqnarray}
\begin{eqnarray}
v^{2nd}_{i}(\overrightarrow{r}) &=& \frac{1}{{T_{gen}}}\frac{1}{{G(\overrightarrow{r},\overrightarrow{r})}}\times\nonumber \\
&& \left [\frac{1}{{2}}\frac{\partial^{2} G(\overrightarrow{r},\overrightarrow{r}')}{{\partial x^{2}_{i}}}\left ( \overline{x^{3}_{i}}-\overline{x_{i}^{2}}\overline{x_{i}} \right )+\sum_{i\neq j}\left ( \overline{x^{2}_{i}}-\overline{x_{i}}^{2} \right )\overline{x_{j}}\left (\frac{\partial G(\overrightarrow{r},\overrightarrow{r}')}{{\partial x_{i}\partial x'_{j}}}+\frac{\partial G(\overrightarrow{r},\overrightarrow{r}')}{{\partial x_{i}\partial x_{j}}}\right )-\right .\nonumber \\
&& -\left .\frac{1}{{G(\overrightarrow{r},\overrightarrow{r})}}\frac{\partial G(\overrightarrow{r},\overrightarrow{r}')}{{\partial x_{i}}}\left ( \frac{\partial G(\overrightarrow{r},\overrightarrow{r}')}{{\partial x_{j}}}+\frac{\partial G(\overrightarrow{r},\overrightarrow{r}')}{{\partial x'_{j}}}\right )\overline{x_{j}}\right ], \label{SMvival2}%
\end{eqnarray}
\end{widetext}
and $\overline{x^{n}_{i}}=\int \rho (x_{i}-x_{0i})^{n}$. For derivation of eqs. (\ref{SMvival1}) and (\ref{SMvival2}), see section \ref{SMsecDvi12}.

The spread $\left ( \overline{x^{2}_{i}}-\overline{x_{i}}^{2} \right )$ and asymmetry $\overline{x_{i}}$ are estimated in the limits of slow and fast propagations of the population in the phenotype space. For derivation of the following equations see section \ref{SMDspsk}. In the limit of almost stable population, the spread is defined by diffusion in the vicinity of the stable position:
\begin{eqnarray}
\left ( \overline{x^{2}_{i}}-\overline{x_{i}}^{2} \right ) &\propto& 2\sqrt{-DT_{gen}\left( \frac{\partial^{2} G(\overrightarrow{r},\overrightarrow{r}')}{{\partial x^{2}_{i}}} \right )^{-1}}
,\nonumber \\ \label{SMspreadsl}
\end{eqnarray}
\begin{eqnarray}
\overline{x_{i}} &\propto&
-\frac{\partial G(\overrightarrow{r},\overrightarrow{r}')}{{\partial x_{i}}}
\left( \frac{\partial^{2} G(\overrightarrow{r},\overrightarrow{r}')}{{\partial x^{2}_{i}}} \right )^{-1}
. \label{SMskewsl}%
\end{eqnarray}
The population remains confined under condition:
\begin{eqnarray}
\frac{\partial^{2} G(\overrightarrow{r},\overrightarrow{r}')}{{\partial x^{2}_{i}}} < 0. \label{SMskew}%
\end{eqnarray}

In case of a fast propagating population, the spread is equal to the diffusion during the time that takes the population to pass its own size:
\begin{eqnarray}
\left ( \overline{x^{2}_{i}}-\overline{x_{i}}^{2} \right ) &\propto& (DT_{gen})^{\frac{2}{{3}}}\left (\frac{\partial G(\overrightarrow{r},\overrightarrow{r}')}{{\partial x_{i}}}\right )^{-\frac{2}{{3}}}, \label{SMspreadfa}
\end{eqnarray}
The corresponding asymmetry of the propagating population is:
\begin{eqnarray}
\overline{x_{i}} &\propto&
(DT_{gen})^{\frac{1}{{3}}}\left (\frac{\partial G(\overrightarrow{r},\overrightarrow{r}')}{{\partial x_{i}}}\right )^{-\frac{1}{{3}}}. \label{SMskewfa}%
\end{eqnarray}

Diversity (multi-stability) of the dynamics of a confined population requires local stable states $\overrightarrow{v}=0$ either since:
\begin{eqnarray}
\frac{\partial G(\overrightarrow{r},\overrightarrow{r}')}{{\partial x_{s}}}= 0, \label{SMstabcondv0}%
\end{eqnarray}
or if further propagation is impossible due to the boundaries of the phenotype space:
\begin{eqnarray}
\overrightarrow{v(\overrightarrow{r_{b}})}\overrightarrow{n} < 0, \label{SMstabcondvperp}%
\end{eqnarray}
where $\overrightarrow{r_{b}}$ is a point on the boundary of the phenotype space and $\overrightarrow{n}$ is the normal vector to the boundary.

In this work, a special case of:
\begin{eqnarray}
\frac{\partial G(\overrightarrow{r},\overrightarrow{r}')}{{\partial x_{s}}}=\frac{\partial^{2} G(\overrightarrow{r},\overrightarrow{r}')}{{\partial x^{2}_{s}}} = 0, \label{SMstabcondDP}%
\end{eqnarray}
is important, in addition to the conditions (\ref{SMstabcondv0}) and (\ref{SMstabcondvperp}).
The corresponding velocity $v_{s}$ depends entirely on the evolution in other directions:
\begin{widetext}
\begin{eqnarray}
v^{1st}_{s}(\overrightarrow{r}) &=& 0\nonumber \\
v^{2nd}_{s}(\overrightarrow{r}) &=& \frac{1}{{T_{gen}}}\frac{1}{{G(\overrightarrow{r},\overrightarrow{r})}}\sum_{j\neq s}\left ( \overline{x^{2}_{s}}-\overline{x_{s}}^{2} \right )\overline{x_{j}}\left (\frac{\partial G(\overrightarrow{r},\overrightarrow{r}')}{{\partial x_{s}\partial x'_{j}}}+\frac{\partial G(\overrightarrow{r},\overrightarrow{r}')}{{\partial x_{s}\partial x_{j}}}\right ).\nonumber \\
\label{SMvi2fin}%
\end{eqnarray}
\end{widetext}
A population can be stable at many different points along $s$, provided that for each value of $s$ there are stable values for other phenotype coordinates (see (\ref{SMstabcondv0})).

The equations for $v^{1st}_{i}(\overrightarrow{r})$, $v^{1st}_{i}(\overrightarrow{r})$, $\left ( \overline{x^{2}_{i}}-\overline{x_{i}}^{2} \right )$ and $\overline{x_{i}}$ define the main properties of the evolutionary dynamics of a confined population; First,
the lowest order contribution $\overrightarrow{v^{1st}}$ to the velocity of a population corresponds to the deviation from stability condition:
\begin{eqnarray}
\frac{\partial G(\overrightarrow{r},\overrightarrow{r}')}{{\partial x_{i}}} = 0, \label{SMstabcondDP1}%
\end{eqnarray}
in case where (\ref{SMstabcond1}) holds.
Second, the motion in one direction contributes to the velocities in other directions; $v^{1st}_{j}\neq 0$ designates asymmetry along the $j$ direction $\overline{x_{j}}\neq 0$ and finite contribution to $v^{2nd}_{i}\neq 0$ for $i\neq j$, see eq. (\ref{SMvival2}).

\subsection{Evolution of information exchange}

The analysis proceeds along the following steps: First, all possible individual strategies (the phenotype space) are defined. Second, a special set of the competitions (tables of payoffs $W_{pq}$) is demonstrated to satisfy the evolutionary stability of the phenotypes, corresponding to the maximum information exchange. Third, mutual affective behavior during a competition and zero payoff for a non-responsive (stochastic) behavior are demonstrated to be essential for reasonable evolutionary dynamics and its diversity (multi-stability). The main result is that fluctuations of evolutionary conditions accelerate the development of information exchange.

\subsubsection{Individual phenotypes}

In this work, a two-reaction phenotype is composed of different behavior modes $k$ ($k\in \{1,N\}$). A specific behavior mode is described by a table of payoffs $W^{k}_{pq}$ (see (\ref{SMWtable1})) and three parameters $(\epsilon_{k},\alpha_{k},\beta_{k})$: an individual possesses the probabilities $\epsilon_{k}$ to be involved in a competition having payoffs $W^{k}_{pq}$, and conditional probabilities $\alpha_{k}$ and $\beta_{k}$ defining its method (strategy) to choose one of the reactions. The behavior modes differ from each other by their payoffs $W^{k}_{pq}$ and by the allowed values of $\alpha$ and $\beta$.

Parameters $\alpha_{k}$ and $\beta_{k}$ define the statistics of reactions of an individual in mode $k$ against selfishness and cooperation correspondingly: $\alpha_{k}$ and $1-\alpha_{k}$ are the conditional probabilities to choose cooperation $C$ and selfish $S$ reactions against the selfish reaction $S$, while $\beta_{k}$ and $1-\beta_{k}$ are the conditional probabilities to choose cooperation $C$ and selfish $S$ reactions against the cooperation reaction $C$.

To take into account affective behavior, additional assumptions are required: First, the parameters $(\epsilon_{k},\alpha_{k},\beta_{k})$ depend on each other, e.g. $\epsilon^{A}_{k}(\epsilon^{B}_{j},\alpha^{B}_{j},\beta^{B}_{j})$ indicates that individual $B$ affects the probability of the individual $A$ to be in behavior mode $k$. Second, $W^{k}_{pq}$ can depend on the probability to be in behavior mode $\epsilon_{k}$ (frequency dependent payoffs), as in the case of an evolutionary reward that is divided either across many competitions or during a major one. The payoffs $W^{k}_{pq}$, however, are assumed to be independent of the probabilities $\alpha$ and $\beta$.

\subsubsection{Interaction of two individuals: payoffs}
\label{SMsecITIP}
To find the evolutionary payoff (\ref{SMpayoff0}) of an individual $m$ in mode $i$ interacting with an individual $h$ in mode $j$, one should calculate the statistics of the mutual reactions $\Omega_{pq}(m_{i},h_{j})$ and mutual behavior modes $\Xi(m_{i},h_{j})$ during a competition, as function of the individual parameters $\epsilon^{m}_{i},\alpha^{m}_{i},\beta^{m}_{i}$ and $\epsilon^{h}_{j},\alpha^{h}_{j},\beta^{h}_{j}$.

According to the definition of $\epsilon$:
\begin{eqnarray}
\Xi(\epsilon_{i},\alpha_{i},\beta_{i}|\epsilon_{j},\alpha_{j},\beta_{j})&=&\epsilon_{i}\epsilon_{j}, \label{SMximh}
\end{eqnarray}
indicating that the probability of a mode $i$ against mode $j$ competition is the product of the individual probabilities to be in the modes $i$ and $j$ correspondingly.

To derive the statistics of the mutual reactions $\Omega_{pq}(m_{i},h_{j})$ as a function of $(\alpha_{i},\beta_{i})$ and $(\alpha_{j},\beta_{j})$, one should solve the self-consistent system of equations:
\begin{eqnarray}
&&\alpha_{i}=\frac{\Omega_{CS}}{{\Omega_{CS}+\Omega_{SS}}},\;\alpha_{j}=\frac{\Omega_{SC}}{{\Omega_{SS}+\Omega_{SC}}},\nonumber \\
&&\beta_{i}=\frac{\Omega_{CC}}{{\Omega_{SC}+\Omega_{CC}}},\;\beta_{j}=\frac{\Omega_{CC}}{{\Omega_{CC}+\Omega_{CS}}},\nonumber \\
&&\Omega_{SS}+\Omega_{SC}+\Omega_{CS}+\Omega_{CC}=1, \label{SMomegasys}%
\end{eqnarray}
following the definitions of $\alpha$ and $\beta$.

Causality makes solution of (\ref{SMomegasys}) in a general case impossible; in the course of pairwise interactions, simultaneous conditional reactions and information exchange are impossible. The second to respond, therefore, possesses an advantage by knowing the reaction of the competitor. Exact solution, however, exists for some cases, turning out to be the most relevant for the evolution of information exchange (see section \ref{SMsecGFDD}).

The solution of the system (\ref{SMomegasys}) exists for the case of $(\alpha_{i},\beta_{i})=(\alpha_{j},\beta_{j})$ and at the boundaries of the $(\alpha_{j},\beta_{j})$ space: $\alpha_{i,j}=0,1$ or $\beta_{i,j}=0,1$ (see section \ref{SMdermat}).
According to the definition of $\alpha$ and $\beta$, the solution of (\ref{SMomegasys}) comes in the form:
\begin{eqnarray}
\Omega_{SS}(\epsilon_{i},\alpha_{i},\beta_{i}|\epsilon_{j},\alpha_{j},\beta_{j})&=&(1-\alpha_{i})\gamma_{ji},\nonumber \\
\Omega_{SC}(\epsilon_{i},\alpha_{i},\beta_{i}|\epsilon_{j},\alpha_{j},\beta_{j})&=&(1-\beta_{i})(1-\gamma_{ji}),\nonumber \\
\Omega_{CS}(\epsilon_{i},\alpha_{i},\beta_{i}|\epsilon_{j},\alpha_{j},\beta_{j})&=&\alpha_{i}\gamma_{ji},\nonumber \\
\Omega_{CC}(\epsilon_{i},\alpha_{i},\beta_{i}|\epsilon_{j},\alpha_{j},\beta_{j})&=&\beta_{i}(1-\gamma_{ji}),
\label{SMomegaAB}%
\end{eqnarray}
where $\gamma_{ji}$ is the probability of individual $j$ to provide reaction $S$ during a competition with the individual $i$. It follows from the self-consistent (mean field) conditions:
\begin{eqnarray}
\gamma_{ji}=(1-\alpha_{j})\gamma_{ij}+(1-\beta_{j})(1-\gamma_{ij}),\nonumber \\
\gamma_{ij}=(1-\alpha_{i})\gamma_{ji}+(1-\beta_{i})(1-\gamma_{ji}), \label{SMgammajisys}%
\end{eqnarray}
indicating that probabilities $\gamma_{ji}$ and $\gamma_{ij}$ correspond to the same values of $\Omega_{pq}(m_{i},h_{j})$. Solving this system of equations results in:
\begin{equation}
\gamma_{ji}=\frac{(1-\beta_{j})-(1-\beta_{i})(\alpha_{j}-\beta_{j})}{{1-(\alpha_{i}-\beta_{i})(\alpha_{j}-\beta_{j})}}. \label{SMgammaji}%
\end{equation}
In a homogeneous population of identical individuals $(\epsilon_{i},\alpha_{i},\beta_{i})$, eq. (\ref{SMgammaji}) becomes:
\begin{equation}
\gamma_{ii}=\frac{1-\beta}{{1+\alpha-\beta}}, \label{SMgammaii}%
\end{equation}
where $\gamma_{ii}$ is independent of $\epsilon_{i}$. For derivation of eq. (\ref{SMgammaii}), see section \ref{SMDgii}.

The total gain of an individual $m$ in an interaction with an individual $h$ (see eq. (\ref{SMpayoff0})), becomes:
\begin{equation}
G(m,h)=\sum_{i,j}\epsilon^{m}_{i}\epsilon^{h}_{j}P(m_{i}|h_{j}|W^{i}_{pq}), \label{SMgainmh}%
\end{equation}
where summation goes over all behavior modes $i,j$.
The payoff for a competition $P(m_{i}|h_{j}|W^{i}_{pq})$ (\ref{SMpayoff1}), taking into account (\ref{SMomegaAB}), becomes:
\begin{widetext}
\begin{eqnarray}
P(m_{i}|h_{j}|W^{i}_{pq}) &=& (1-\alpha^{m}_{i})\gamma_{ji}W^{i}_{SS}+(1-\beta^{m}_{i})(1-\gamma_{ji})W^{i}_{SC}+\ldots\nonumber \\
&&+\alpha^{m}_{i}\gamma_{ji}W^{i}_{CS}+\beta^{m}_{i}(1-\gamma_{ji})W^{i}_{CC}, \label{SMpayoffmh}%
\end{eqnarray}
\end{widetext}
where $\gamma_{ji}$ is defined by eq. (\ref{SMgammaji}).

In the general case, $\Omega_{pq}$ can be derived as a function of $(\alpha_{i},\beta_{i})$, $(\alpha_{j},\beta_{j})$, $\gamma_{ij}$, $\gamma_{ji}$ and additional parameters describing causality issues in the course of pairwise interaction. The arrangement to be the first or the second to respond during a competition can be defined in many different ways; e.g. it can be random or an individual with more advanced abilities to process information can possess an advantage. For instance, in case of equiprobable right to be the second to respond, the statistics of the mutual reactions is:
\begin{equation}
\Omega^{\ast}_{pq}(i|j)=\frac{\Omega_{pq}(i|j)+\Omega_{pq}(j|i)}{{2}}, \label{SMomegaeqi}%
\end{equation}
where $\Omega_{pq}(i|j)$ corresponds to (\ref{SMomegaAB}). Fortunately, the main conclusions of this work are independent of the specific choice of the causality mechanism.

\subsubsection{Information exchange of two identical individuals}
\label{SMsubsecIETII}
An amount of information must be transferred between two interacting strangers to allow correlated, rather than random, mutual reactions. Information exchange between two interacting individuals follows from the statistics of the mutual reactions $\Omega_{pq}$:
\begin{eqnarray}
I=\sum_{p,q=S,C}\left [\Omega_{pq}\log_{2}\left (\frac{\Omega_{pq}}{{(\Omega_{pS}+\Omega_{pC})(\Omega_{Sq}+\Omega_{Cq})}}\right )\right ],\nonumber \\ \label{SMmutinfo1}
\end{eqnarray}
under assumption that no information is shared prior to the interaction.

To derive information exchange per behavior mode, one should substitute eqs. (\ref{SMgammaii}) and (\ref{SMomegaAB}) into eq. (\ref{SMmutinfo1}), assuming that the individuals compete in the same mode $(\alpha_{i},\beta_{i})=(\alpha_{j},\beta_{j})=(\alpha,\beta)$. The corresponding statistics of mutual reactions (\ref{SMomegaAB}) is:
\begin{eqnarray}
\Omega_{SS}=(1-\alpha)\gamma,\;\Omega_{CC}=\beta(1-\gamma) \nonumber \\
\Omega_{CS}=\alpha\gamma,\;\Omega_{SC}=(1-\beta)(1-\gamma),
\label{SMomegaABeq}%
\end{eqnarray}
where, according to eq. (\ref{SMgammaii}) $\gamma$ is:
\begin{eqnarray}
\gamma=\frac{1-\beta}{{1+\alpha-\beta}}, \label{SMgammajieq}%
\end{eqnarray}
and index $i$ is omitted as unnecessary in this case.

Substitution of eqs. (\ref{SMomegaABeq}) and (\ref{SMgammajieq}) into eq. (\ref{SMmutinfo1}) results in:
\begin{widetext}
\begin{eqnarray}
I&=&(1-\alpha)\gamma\log_{2}\left (\frac{(1-\alpha)\gamma}{{\gamma^{2}}} \right )+(1-\beta)(1-\gamma)\log_{2}\left (\frac{(1-\beta)(1-\gamma)}{{\gamma(1-\gamma)}} \right)+\ldots\nonumber \\
&&+\alpha\gamma\log_{2}\left (\frac{\alpha\gamma}{{(1-\gamma)\gamma}} \right)+\beta(1-\gamma)\log_{2}\left (\frac{\beta(1-\gamma)}{{(1-\gamma)^{2}}} \right),
\label{SMmutinfo2expand}
\end{eqnarray}
\end{widetext}
taking into account that for $p,q=S$:
\begin{eqnarray}
\Omega_{pS}+\Omega_{pC}=\Omega_{Sq}+\Omega_{Cq}=\gamma,
\label{SMomegasumcond1}%
\end{eqnarray}
and, in case of $p,q=C$
\begin{eqnarray}
\Omega_{pS}+\Omega_{pC}=\Omega_{Sq}+\Omega_{Cq}=1-\gamma,
\label{SMomegasumcond1}%
\end{eqnarray}
since these summations correspond to the unconditional probabilities to provide either selfish or cooperative reactions (see Fig. 2B).

The expression (\ref{SMmutinfo2expand}) can be reduced by symbolic manipulations to:
\begin{eqnarray}
I=\log_{2}\left (\frac{(1-\alpha)^{(1-\alpha)\gamma}(1-\beta)^{(1-\beta)(1-\gamma)}\alpha^{\alpha\gamma}\beta^{\beta(1-\gamma)}}{{\gamma^{\gamma}(1-\gamma)^{\gamma}}}\right ),\nonumber \\
\label{SMmutinfo2}
\end{eqnarray}
where $\gamma$ corresponds to (\ref{SMgammajieq}).

The main properties of information exchange (\ref{SMmutinfo2}) are presented in Fig. 2C. $I$ varies from $0$ to $1$ bit, describing a binary reaction. The maximal transfer ($1$ bit per interaction) corresponds to the indirect reciprocity $(IR)$ $(\alpha=1,\beta=0)$ and synchronous $(\alpha=0,\beta=1)$ behavior modes. No information transfer is required in cases of random reactions $(\alpha=\beta)$, homogeneous cooperation $(\beta=1)$ and selfishness $(\alpha=0)$.

\subsubsection{Instability of the indirect reciprocity and extremely slow development of synchronization in populations with a single behavior mode}

The main result of this section is that at least two behavior modes are required to evolve significant information exchange. In populations composed of individuals with a single behavior mode $(\alpha,\beta)$, indirect reciprocity $(\alpha=1,\beta=0)$ is always unstable and an infinite time is required to develop the synchronization $(\alpha=0,\beta=1)$.

A population composed of a host with a single indirect reciprocity behavior mode $(\alpha=1,\beta=0)$, is always unstable either toward $\alpha$ $(\alpha=1-\Delta\alpha,\beta=0)$ or $\beta$ $(\alpha=1,\beta=\Delta\beta)$ mutants. An individual $(\alpha_{i}=1,\beta_{i}=0)$ expresses unconditional selfishness $\gamma_{ji}=0$ or cooperation $\gamma_{ji}=1$ against mutants $(\alpha_{j}=1-\Delta\alpha,\beta_{j}=0)$ and $(\alpha_{j}=1,\beta_{j}=\Delta\beta)$ $(\Delta\alpha,\Delta\beta\ll0)$ correspondingly, though possesses $\gamma_{ii}=0.5$ for interaction with an identical individual:
\begin{eqnarray}
\gamma_{ji}(\alpha=1,\beta=0|\alpha=1,\beta=0)=\gamma_{ii}(1,0)=\frac{1}{{2}}, \label{SMgammalimeq}
\end{eqnarray}
\begin{eqnarray}
\lim_{\Delta\alpha \to 0}\gamma_{ji}(\alpha=1,\beta=0|\alpha=1-\Delta\alpha,\beta=0) = 0 \label{SMgammalimalpha}
\end{eqnarray}
\begin{eqnarray}
\lim_{\Delta\beta \to 0}\gamma_{ji}(\alpha=1,\beta=0|\alpha=1,\beta=\Delta\beta) = 1,\label{SMgammalimbeta}
\end{eqnarray}
The corresponding payoffs (\ref{SMpayoff1}) are (taking into account eqs. (\ref{SMgammalimeq}),(\ref{SMgammalimalpha}),(\ref{SMgammalimbeta}) and (\ref{SMomegaAB})):
\begin{eqnarray}
G(h,h)=(W_{SC}+W_{CS})/2, \label{SMPhh}
\end{eqnarray}
\begin{eqnarray}
G(m_{\alpha},h)=W_{SC}, \label{SMPmah}
\end{eqnarray}
\begin{eqnarray}
G(m_{\beta},h)=W_{CS},\label{SMPmbh}
\end{eqnarray}
One of the mutants, therefore, possesses a payoff greater than the host, since either:
\begin{eqnarray}
W_{SC}>(W_{SC}+W_{CS})/2,\label{SMW12sum}
\end{eqnarray}
or:
\begin{eqnarray}
W_{CS}>(W_{SC}+W_{CS})/2.\label{SMW21sum}
\end{eqnarray}
Consequently, indirect reciprocity can not develop in populations with a single behavior mode, since the condition (\ref{SMstabcond1}) is always broken.

The stability of synchronous population is ill-defined: it can be stable, though requires infinite time to develop. Populations with $\alpha=0$ are identical with homogeneous reactions $S$ vs. $S$, while populations with $\beta=1$ are identical with homogeneous reactions $C$ vs. $C$. The dynamics near these axes are very slow, since there is no evolutionary benefit for transfer between two identical states. A population, therefore, can be stacked in the vicinity of the synchronous mode $(\alpha=0,\beta=1)$ but is unable to reach it.

\subsubsection{Stability and diversity of indirect reciprocity}
\label{SMsecSDIR}
Consider a population composed of individuals, possessing several behavior modes $(\epsilon_{k},\alpha_{k},\beta_{k})$ ($k=1..N$). The main goal is to find the evolutionary conditions (payoffs $W^{k}_{pq}$, mechanisms of affective behavior and constraints on the exchange of information) required for the development and stability of the indirect reciprocity mode $(\alpha_{IR}=1,\beta_{IR}=0)$. An additional requirement is the diversity of the evolutionary dynamics, meaning that multiple stable states of a population exist under the same evolutionary conditions.

We will demonstrate that stability of the indirect reciprocity requires greater evolutionary payoff for the less frequent pairwise interactions. The individual payoff $W^{IR}_{pq}$ for a pairwise interaction, must be inversely proportional to $\epsilon_{IR}$, the probability of the individual to be found in the indirect reciprocity mode:
\begin{eqnarray}
W^{IR}_{pq}\propto\frac{1}{{\epsilon_{IR}}}.\label{SMmestabcond}
\end{eqnarray}
In this case, the table of payoffs (\ref{SMWtable1}) is:
\begin{equation}
\frac{1}{{\epsilon_{mode}}}\times
\begin{tabular}{c|c|c}
  & $C$ & $S$ \\
\hline
$C$ & $W^{mode}_{CC}$ & $W^{mode}_{CS}$ \\
\hline
$S$ & $W^{mode}_{SC}$ & $W^{mode}_{SS}$  \\
\end{tabular}, \label{SMWtable2}
\end{equation}
where $W^{mode}_{pq}$ corresponds to the possible evolutionary payoff for a such behavioral mode.

Condition (\ref{SMmestabcond}) is essential for evolutionary stability of the indirect reciprocity mode $(\alpha_{1}^{h}=1,\beta_{1}^{h}=0)$ in a population $(\epsilon_{k}^{h},\alpha_{k}^{h},\beta_{k}^{h})$ against either $(\alpha_{1}^{m}=1-\Delta\alpha,\beta_{1}^{m}=0)$ or $(\alpha_{1}^{m}=1,\beta_{1}^{m}=\Delta\beta)$ mutants, $\Delta\alpha,\Delta\beta\ll 1$. The corresponding conditions of the evolutionary stability (\ref{SMstabcond1}) are (see section \ref{SMDabstab}), for the case of $(\alpha_{1}^{m}=1-\Delta\alpha,\beta_{1}^{m}=0)$:
\begin{equation}
\epsilon_{1}^2\frac{W^{1}_{CS}-W^{1}_{SC}}{{2}}+\Delta\alpha\epsilon_{1}B>0, \label{SMmestabcondalpha}
\end{equation}
and for the case $(\alpha_{1}^{m}=1,\beta_{1}^{m}=-\Delta\beta)$:
\begin{equation}
\epsilon_{1}^2\frac{W^{1}_{SC}-W^{1}_{CS}}{{2}}+\Delta\beta\epsilon_{1}D>0, \label{SMmestabcondbeta}
\end{equation}
under assumption that $\epsilon_{1}^{h}=\epsilon_{1}^{m}=\epsilon_{1}$ and keeping only the first order terms of $\Delta\alpha$ and $\Delta\beta$. Inequalities (\ref{SMmestabcondalpha}) and (\ref{SMmestabcondbeta}) can not hold simultaneously without condition (\ref{SMmestabcond}); the terms $\propto\epsilon_{1}^2$ are dominant and possess opposite signs.

Under condition (\ref{SMmestabcond}), the expressions (\ref{SMmestabcondalpha}) and (\ref{SMmestabcondbeta}) become:
\begin{equation}
\epsilon_{1}\frac{W^{1}_{CS}-W^{1}_{SC}}{{2}}+\Delta\alpha \tilde{B}>0, \label{SMmestred1}
\end{equation}
\begin{equation}
\epsilon_{1}\frac{W^{1}_{SC}-W^{1}_{CS}}{{2}}+\Delta\beta \tilde{D}>0, \label{SMmestred2}
\end{equation}
taking into account that both $B$ and $D$ are proportional to $W^{1}_{pq}$ (see section \ref{SMDabstab}). In the case $\epsilon_{1}\rightarrow 0$, (\ref{SMmestred1}) and (\ref{SMmestred2}) converge to $\Delta\alpha \tilde{B}$ and $\Delta\beta \tilde{D}$ correspondingly. The stability of the indirect reciprocity behavior mode is, therefore, possible when the probability to be in such a mode is low $(\epsilon_{IR}\rightarrow 0,\alpha_{IR}=1,\beta_{IR}=0)$ and $\tilde{B},\tilde{D}>0$. A scaling of the payoffs $W_{pq}\propto\epsilon^{\lambda}$, $\lambda\neq -1$ (different from (\ref{SMmestabcond})) either prevents stability of the indirect reciprocity mode $(\lambda>-1)$ or makes the expression for evolutionary gain (\ref{SMgainmh}) to diverge $(\lambda<-1)$.

Stability analysis of the indirect reciprocity mode $(\epsilon_{IR}\rightarrow 0,\alpha_{IR}=1,\beta_{IR}=0)$ against $(\epsilon_{1}=1-\Delta\epsilon,\alpha_{1}^{m}=1,\beta_{1}^{m}=0)$ mutants, brings no significant confinements on the required evolutionary conditions; this mode is stable for changes in $\epsilon$ under quite general choice of payoffs.

Affective behavior is required to allow for the development of the indirect reciprocity behavior mode $(\epsilon_{IR}\rightarrow 0,\alpha_{IR}=1,\beta_{IR}=0)$. The condition (\ref{SMmestabcond}) suggests two different interpretations of the parameter $\epsilon_{i}$, either as an individual probability to be in mode $i$ or as a probability to affect the competitor to behave in mode $i$. The corresponding payoff for an individual $(\alpha_{i},\beta_{i},\epsilon_{i})$ competing against $(\alpha_{j},\beta_{j},\epsilon_{j})$ is:
\begin{equation}
G(A,B)=\sum_{ij}\epsilon^{B}_{j}P(\alpha^{A}_{i},\beta^{A}_{i}|\alpha^{B}_{j},\beta^{B}_{j}|W^{i}_{pq}),
\label{SMpab}
\end{equation}
or, in case of affective behavior:
\begin{equation}
G(A,B)=\sum_{ij}\epsilon^{A}_{i}P(\alpha^{A}_{i},\beta^{A}_{i}|\alpha^{B}_{j},\beta^{B}_{j}|W^{i}_{pq}),
\label{SMpabaff}
\end{equation}
In the limit of $\epsilon^{h}=\epsilon^{m}$ these interpretations are indistinguishable. The development of $\epsilon_{IR}\rightarrow 0$, required for the stability of the indirect reciprocity, is impossible in case of (\ref{SMpab}); $v_{\epsilon}=0$, since derivatives for any order $n$ vanish, $\partial^{n}G(\overrightarrow{r_{i}},\overrightarrow{r_{j}})/\partial \epsilon_{i}^{n}=0$ (see (\ref{SMvival1}) and(\ref{SMvival2})). Consequently, affective behavior (\ref{SMpabaff}) is essential to ensure evolutionary dynamics of a confined population subjected to condition (\ref{SMmestabcond}).

Diversity requires the existence of at least two different modes (e.g. mode $1$ and mode $2$), that in interaction with an arbitrary mode $i$ will provide an equal payoff:
\begin{eqnarray}
P(\alpha_{i},\beta_{i}|\alpha_{1},\beta_{1}|W^{i}_{pq})=P(\alpha_{i},\beta_{i}|\alpha_{2},\beta_{2}|W^{i}_{pq}).
\label{SMCeq}
\end{eqnarray}
The multi-stability along $\epsilon_{1}$ direction requires:
\begin{equation}
\frac{\partial G(m,h)}{{\partial \epsilon^{m}_{1}}} = 0, \label{SMdivcond1}
\end{equation}
for multiple values of $\epsilon_{1}$. Diversity along $\alpha$ and $\beta$ directions does not occur unless the weights $W_{pq}$ depend on $\alpha$ and $\beta$. Condition (\ref{SMCeq}) follows from eq. (\ref{SMpabaff}) and $\epsilon_{2}=1-\sum_{j\neq 2} \epsilon_{j}$:
\begin{equation}
\frac{\partial G(m,h)}{{\partial \epsilon^{m}_{1}}} = \sum_{i}(P(\alpha_{i},\beta_{i}|\alpha_{1},\beta_{1}|W^{i}_{pq})-P(\alpha_{i},\beta_{i}|\alpha_{2},\beta_{2}|W^{i}_{pq})), \label{SMdivcond2}
\end{equation}
taking into account (\ref{SMdivcond1}) and assuming $W^{i}_{pq}$ to be independent for different $i$. The modes $1$ and $2$ have to be linked by some permanent constraint to ensure condition (\ref{SMCeq}).

The condition for diversity (\ref{SMCeq}) holds if for any $i$, either $\gamma_{1i}=\gamma_{2i}$ (assuming the payoffs $P$ defined by (\ref{SMpayoffmh})) or $W^{i}_{pq}=0$. For instance, in case of only two behavior modes: responsive $(R)$ and non-responsive $(NR)$, the non-responsive mode may correspond to stochastic behavior with reaction statistics $\gamma_{R}$ matching the reaction statistics of the individual in mode $(R)$:
\begin{eqnarray}
\alpha_{NR}=\beta_{NR}=1-\gamma_{R},\label{SMdivcond3}
\end{eqnarray}
leading to:
\begin{eqnarray}
P(\alpha_{R},\beta_{R}|\alpha_{R},\beta_{R}|W^{R}_{pq})-\ldots\nonumber \\ -P(\alpha_{R},\beta_{R}|\alpha_{NR},\beta_{NR}|W^{R}_{pq})=0.\label{SMPdpR}
\end{eqnarray}
Zero payoff for non-responsive behavior:
\begin{eqnarray}
W^{NR}_{pq}=0,\label{SMdivcond4}
\end{eqnarray}
is required to ensure:
\begin{eqnarray}
P(\alpha_{NR},\beta_{NR}|\alpha_{R},\beta_{R}|W^{NR}_{pq})-\ldots\nonumber \\ -P(\alpha_{NR},\beta_{NR}|\alpha_{NR},\beta_{NR}|W^{NR}_{pq})=0,\label{SMPdpNR}
\end{eqnarray}
since the payoffs can not be the same for an interactions with (responsive vs. non-responsive) and without (non-responsive vs. non-responsive) information transfer. Diversity is preserved in the case of many behavior modes under condition that an individual's non-responsive behavior depends on the competitor's mode $i$, matching the average statistics of reactions against this specific behavior mode.

The condition (\ref{SMdivcond3}) possesses a reasonable interpretation and brings no additional parameters to the model, contrary to other possibilities to explain the required indistinguishability of the two different modes (\ref{SMCeq}). The non-responsive behavior can be associated with reactions induced by an arbitrary member of the population, rather than by the opponent. In this case the probability of a specific reaction matches the average reaction statistics of the individual.

\subsection{Evolution of information exchange in a population with two behavior modes}
\label{SMsecECAP}
Let us consider an example of a population composed of individuals with two behavior modes: non-responsive ($R$) and responsive ($NR$), and subjected to the conditions (\ref{SMmestabcond}), (\ref{SMCeq}) and (\ref{SMdivcond3}). The corresponding individual phenotype consists of three parameters: selfishness aversion $\alpha$, cooperation attraction $\beta$ and ability of affect $\epsilon$. In the responsive mode, $\alpha$ and $1-\alpha$ are the probabilities to choose cooperation $C$ and selfish $S$ reactions against the competitor's selfish reaction $S$, while $\beta$ and $1-\beta$ are the probabilities to choose cooperation $C$ and selfish $S$ reactions against the competitor's cooperation reaction $C$. In the non-responsive mode it generates selfish and cooperative reactions with probabilities $\gamma$ and $1-\gamma$ correspondingly, where $\gamma$ is the individual probability to be in state $S$ averaged over all (both non-responsive and responsive) interactions. Ability of affect $\epsilon$ is the probability to put a competitor into its non-responsive mode.

The evolutionary payoffs for each behavior mode are described by the tables of payoffs (\ref{SMWtable1}). There is no payoff for non-responsive behavior $W^{NR}=0$. The weight table for responsive mode $W^{R}$ can be reduced to a two parameter form (see section \ref{SMDtab}):
\begin{equation}
\begin{tabular}{c|c|c}
  & $C$ & $S$ \\
\hline
$C$ & $1$ & $c$ \\
\hline
$S$ & $b$ & $0$  \\
\end{tabular}, \label{SMWtable3}
\end{equation}
when the average fitness in the population is:
\begin{eqnarray}
\bar F\approx 1.
\label{SMfiteq1}
\end{eqnarray}
This condition corresponds to a population that is almost stable in size, since the fitness is a measure of the progeny.

The fluctuations of payoffs $b$ and $c$ correspond to the short time fluctuations in the size of a population $S$:
\begin{eqnarray}
\overline{\Delta c^2},\overline{\Delta b^2}\propto \overline{\left (\frac{\Delta S}{{S}}\right )^2}.
\label{SMbcs}
\end{eqnarray}
In case of non-zero sum game the population size changes with time, and the fluctuations are, therefore, relative to the average growth rate of the population.

\subsubsection{Dynamics}

This section demonstrates that development of information exchange requires payoffs $b$ and $c$ to fluctuate. Otherwise, the population converges to one of its stable points with affect ability $\epsilon\neq 0$, far from the indirect reciprocity mode. The motion of the population, as a consequence of these fluctuations, makes the development of the affect ability $\epsilon\rightarrow 1$ and stability of the indirect reciprocity mode $(\epsilon_{IR}\rightarrow 1,\alpha_{IR}=1,\beta_{IR}=0)$ possible. The evolution of information exchange occurs in the vicinity of the stable points, in case the fluctuations of $b$ and $c$ are small.

Consider a population composed of phenotypes $(\epsilon_{i},\alpha_{i},\beta_{i})$, with corresponding densities $\rho_{i}(\epsilon_{i},\alpha_{i},\beta_{i})$. The payoff (\ref{SMpabaff}) for an individual $i$ interacting with an individual $j$ is:
\begin{eqnarray}
G(i,j) &=& (1-\epsilon_{i})P(\alpha_{i},\beta_{i}|\gamma_{ji}|W^{i}_{pq})+\nonumber \\
&&+\epsilon_{i}P(\alpha_{i},\beta_{i}|\gamma_{j}|W^{i}_{pq}), \label{SMdinpay}%
\end{eqnarray}
where $\gamma_{ji}$ (\ref{SMgammaji}) and $\gamma_{j}$ are the probabilities of individual $j$ to provide reaction $S$ in its responsive and non-responsive behavior modes correspondingly. The payoff $P$ according to eq. (\ref{SMpayoffmh}) becomes:
\begin{equation}
P(\alpha,\beta|\gamma|W_{pq})=\alpha\gamma c+(1-\beta)(1-\gamma)b+\beta(1-\gamma), \label{SMpayoftbbc}%
\end{equation}
taking (\ref{SMWtable3}) into account.

The probability $\gamma_{i}$ follows from the self-consistent system of equations similar to eqs. (\ref{SMgammajisys}):
\begin{widetext}
\begin{eqnarray}
\gamma_{i}=\sum_{j} \left[(1-\epsilon_{j})\rho_{j}((1-\alpha_{i})\gamma_{ji}+(1-\beta_{i})(1-\gamma_{ji}))+\epsilon_{j}\rho_{j}((1-\alpha_{i})\gamma_{j} +(1-\beta_{i})(1-\gamma_{j})\right],\nonumber \\
\label{SMrealsys}%
\end{eqnarray}
\end{widetext}
averaging individual probability to be in state $S$ over all pairwise interactions. In a homogeneous population $(\epsilon,\alpha,\beta)$:
\begin{eqnarray}
\gamma_{j}= \gamma_{ji} = \frac{1-\beta}{{1+\alpha-\beta}}, \label{SMgigjigii}%
\end{eqnarray}
indicating that the non-responsive and responsive behaviors are indistinguishable from each other. For derivation of eq. (\ref{SMgigjigii}), see section \ref{SMDgii}.

The motion of a confined population is derived by substituting the individual gain (\ref{SMdinpay}) into the equations for the velocity of the population (\ref{SMvival1}) and (\ref{SMvi2fin}). The first order $v^{1st}_{\epsilon}$ vanishes:
\begin{eqnarray}
v^{1st}_{\epsilon}(\epsilon,\alpha,\beta)\propto P(\alpha_{i},\beta_{i}|\gamma_{j}|W^{i}_{pq})-P(\alpha_{i},\beta_{i}|\gamma_{ji}|W^{i}_{pq})=0,\nonumber \\
\label{SMvfindyneps1}
\end{eqnarray}
since $\gamma_{j}= \gamma_{ji}$ in this limit (\ref{SMgigjigii}). The local velocity of a population $\overrightarrow{v}=(v^{1st}_{\alpha},v^{1st}_{\beta},v^{2nd}_{\epsilon})$, is defined by:
\begin{eqnarray}
v^{1st}_{\alpha}(\epsilon,\alpha,\beta) &=& \frac{1}{{T_{gen}}}(\overline{x^{2}_{\alpha}}-\overline{x_{\alpha}}^{2})\times\nonumber \\
&&\left [\frac{\partial P(i|j)}{{\partial\alpha_{i}}}+(1-\epsilon_{i})\frac{\partial P(i|j)}{{\partial\gamma_{ji}}}\frac{\partial \gamma_{ji}}{{\partial\alpha_{i}}}\right ],\nonumber \\
\label{SMvfindynalp}
\end{eqnarray}
\begin{eqnarray}
v^{1st}_{\beta}(\epsilon,\alpha,\beta) &=& \frac{1}{{T_{gen}}}(\overline{x^{2}_{\beta}}-\overline{x_{\beta}}^{2})\times\nonumber \\
&&\left [\frac{\partial P(i|j)}{{\partial\beta_{i}}}+(1-\epsilon_{i})\frac{\partial P(i|j)}{{\partial\gamma_{ji}}}\frac{\partial \gamma_{ji}}{{\partial\beta_{i}}}\right ],\nonumber \\
\label{SMvfindynbet}
\end{eqnarray}
\begin{eqnarray}
v^{2nd}_{\epsilon}(\epsilon,\alpha,\beta) &=& -\frac{1}{{T_{gen}}}(\overline{x^{2}_{\epsilon}}-\overline{x_{\epsilon}}^{2})\frac{\partial P(i|j)}{{\partial\gamma_{ji}}}\times\nonumber \\
&&\left [\frac{\partial \gamma_{ji}}{{\partial\alpha_{i}}}\overline{x_{\alpha}}+\frac{\partial \gamma_{ji}}{{\partial\alpha_{j}}}\overline{x_{\alpha}}+ \right .\nonumber \\
&&\left . +\frac{\partial \gamma_{ji}}{{\partial\beta_{i}}}\overline{x_{\beta}}+\frac{\partial \gamma_{ji}}{{\partial\beta_{j}}}\overline{x_{\beta}}\right ]
, \label{SMvi2finab}
\end{eqnarray}
where all derivatives are taken at the point $(\epsilon,\alpha,\beta)=(\epsilon_{i},\alpha_{i},\beta_{i})=(\epsilon_{j},\alpha_{j},\beta_{j})$. The velocity is proportional to the spreads $(\overline{x^{2}_{k}}-\overline{x_{k}^{2}})$ and to asymmetries $\overline{x_{\beta}}$ $(k=\epsilon,\alpha,\beta)$ of the population (see section \ref{EPP}). For derivation of eqs. (\ref{SMvfindyneps1}), (\ref{SMvfindynalp}), (\ref{SMvfindynbet}) and (\ref{SMvi2finab}), see section \ref{SMDabe}.

The payoff $P(i|j)$ is defined by eq. (\ref{SMpayoffmh}) only at the boundaries $(\alpha=0,1)$ or $(\beta=0,1)$. In other regions of $(\alpha,\beta)$ the space eqs. (\ref{SMomegaAB}) are invalid; to calculate the statistics of the mutual reactions $\Omega_{pq}$, the exact order of reactions in a course of the interaction must be taken into account, e.g see (\ref{SMomegaeqi}). The velocities $(v_{\alpha},v_{\beta},v_{\epsilon})$ for arbitrary $(\alpha,\beta)$ are calculated by substituting the payoff $P(i|j)$ in its general form (\ref{SMpayoff1}) into eqs. (\ref{SMvival1}) and (\ref{SMvi2fin}).

\subsection{Evolutionary stable states on the boundaries}
\label{SMessb}

A confined population on the boundary of the $(\epsilon,\alpha,\beta)$ space is evolutionary stable in case of:
\begin{eqnarray}
\left | v_{\parallel}\right | &=& 0, \label{SMstabboundpar}
\end{eqnarray}
\begin{eqnarray}
\frac{\partial v_{\parallel}}{{\partial x_{\parallel}}} &<& 0, \label{SMstabboundsec}
\end{eqnarray}
\begin{eqnarray}
\overrightarrow{v_{\perp}}\overrightarrow{n}&<&0, \label{SMstabboundper}
\end{eqnarray}
where $(v_{\parallel},v_{\perp})$ are the components of the velocity parallel and perpendicular to the boundary, while $\overrightarrow{n}$ is the normal vector at the boundary. The position of a stable point $(\alpha_{st},\beta_{st})$, defined by conditions (\ref{SMstabboundpar}) and (\ref{SMstabboundsec}), is independent of the order of reactions in the course of an interaction, since there exists a solution for (\ref{SMomegaAB}) at the boundary of the phenotype space. To analyze (\ref{SMstabboundper}) the equiprobable order of reactions is assumed (\ref{SMomegaeqi}).

The evolutionary stable points for payoffs $b > 0$ and $b > c$, following (\ref{SMstabboundpar}) and (\ref{SMvfindynalp}), are (see section \ref{SMlast}):
\begin{eqnarray}
\alpha_{st} &=& \frac{-(1-\epsilon)+\sqrt{(1-\epsilon)^2+4\epsilon K^2}}{{2\epsilon K}},\nonumber \\
\beta_{st} &=& 0, \label{SMalphastab}
\end{eqnarray}
where $K=c/b$.

For payoffs $b > 1$ and $b < c$ one obtains, following (\ref{SMstabboundpar}) and (\ref{SMvfindynbet}), the stable points:
\begin{widetext}
\begin{eqnarray}
\alpha_{st} &=& 0,\nonumber \\
\beta_{st} &=& \frac{-(-2+(1-\epsilon)(1-Z))-\sqrt{(-2+(1-\epsilon)(1-Z))^2-4\epsilon(1-\epsilon)Z}}{{2\epsilon}},\nonumber \\
\label{SMbetastab}
\end{eqnarray}
\end{widetext}
where $Z=(c-1)/(b-1)-1$.

For other payoffs $(b,c)$, a population is stable at some arbitrary point on either $\alpha=0$ or $\beta=1$ plane. The evolutionary dynamics in the vicinity of these planes is slow and the stability position is undefined, since $\left | v_{\parallel}\right | = 0$ holds for all points.

In case of developed affect abilities $\epsilon\rightarrow 1$, the indirect reciprocity behavior mode is stable for the region $(b>1,c>0)$, including Chicken, Leader and Battle of the Sexes games. Prisoner's Dilemma degrades information exchange. These results are illustrated in Fig. 4.

\subsection{Games favoring development and degradation of information exchange.}
\label{SMsecGFDD}
The individual ability to exchange information increases with the development of affect ability $\epsilon$. Stability of the indirect reciprocity behavior mode is achieved only at $\epsilon\rightarrow 1$. Consequently, the evolutionary conditions favoring and degrading information exchange correspond to the positive and negative values of $v_{\epsilon}$ (\ref{SMvi2finab}).

Small fluctuations of the payoffs $b$ and $c$ induce motion in the vicinity of the stable points, with the average $\overline{v_{\epsilon}}\propto \overline{\Delta c^{2}},\overline{\Delta b^{2}}$:
\begin{eqnarray}
\overline{v_{\epsilon}}=\left .\frac{1}{{2}}\frac{\partial^{2} v_{\epsilon}(\epsilon,\alpha,\beta,b,c)}{{\partial c^{2}}}\right |_{\alpha_{st},\beta_{st},b,c}\overline{\Delta c^{2}}+\nonumber \\
\left. \frac{1}{{2}}\frac{\partial^{2} v_{\epsilon}(\epsilon,\alpha,\beta,b,c)}{{\partial b^{2}}}\right|_{\alpha_{st},\beta_{st},b,c}\overline{\Delta b^{2}}, \label{SMvedevav}
\end{eqnarray}
under assumption that $\overline{\Delta b},\overline{\Delta c}=0$. In case of constant payoffs $b$ and $c$, a population converges to a stable point with $v_{\epsilon}=0$ (see (\ref{SMvi2finab})).

Information exchange is favored $(\overline{v_{\epsilon}}>0)$ in the Chicken, Battle of the Sexes and Leader games ($(b>0,c<b)$ and $(b>1,c>b)$). The evolutionary dynamics, therefore, occurs in the vicinity of the evolutionary stable points on the axis $\beta=0$ or $\alpha=1$. For the $(b>0,c<b)$ region, the development rate $\overline{v_{\epsilon}}$ follows from eqs. (\ref{SMvedevav}) and (\ref{SMvi2finab}):
\begin{eqnarray}
&&\overline{v_{\epsilon}}(\epsilon,\alpha,\beta) = -\frac{\overline{\Delta p^{2}}}{{2T_{gen}}}(\overline{x^{2}_{\epsilon}}-\overline{x_{\epsilon}}^{2})\left [\frac{\partial^{2}}{{\partial c^{2}}}+\frac{\partial^{2}}{{\partial b^{2}}} \right]\times\nonumber \\
&&\frac{\partial G(i,j)}{{\partial \alpha_{i}}}\left (\frac{\partial^{2} G(i,j)}{{\partial \alpha^{2}_{i}}}\right )^{-1}\frac{\partial P(i|j)}{{\partial\gamma_{ji}}}\left (\frac{\partial \gamma_{ji}}{{\partial\alpha_{i}}}+\frac{\partial \gamma_{ji}}{{\partial\alpha_{j}}}\right )
,\nonumber \\
\label{SMvi2finalp}%
\end{eqnarray}
and for $(b>1,c>b)$ one obtains:
\begin{eqnarray}
&&\overline{v_{\epsilon}}(\epsilon,\alpha,\beta)= -\frac{\overline{\Delta p^{2}}}{{2T_{gen}}}(\overline{x^{2}_{\epsilon}}-\overline{x_{\epsilon}}^{2})\times\nonumber \\
&&\frac{\partial G(i,j)}{{\partial \beta_{i}}}\left (\frac{\partial^{2} G(i,j)}{{\partial \beta^{2}_{i}}}\right )^{-1}\frac{\partial P(i|j)}{{\partial\gamma_{ji}}}\left (\frac{\partial \gamma_{ji}}{{\partial\beta_{i}}}+\frac{\partial \gamma_{ji}}{{\partial\beta_{j}}}\right )
,\nonumber \\
\label{SMvi2finbet}%
\end{eqnarray}
where $\overline{\Delta p^{2}}\approx\overline{\Delta b^{2}},\overline{\Delta c^{2}}$. The asymmetries of a population $\overline{x_{i}}$ are estimated along the boundaries according to (\ref{SMskewsl}) and in the perpendicular direction according to (\ref{SMskewfa}). The asymmetries $\overline{x_{\perp}}$ are neglected since $(DT_{gen})^{\frac{1}{{3}}}\ll 1$, in the limit of small $D$.

The size of the of the population $(\overline{x^{2}_{\epsilon}}-\overline{x_{\epsilon}}^{2})$ along the $\epsilon$ direction is estimated as a diffusion distance during a time $T_{k}$ that takes for a population to pass its own size along the $\alpha$ or $\beta$ directions. It results in:
\begin{eqnarray}
(\overline{x^{2}_{\epsilon}}-\overline{x_{\epsilon}}^{2})\propto \left ( DT_{gen}\right )^{\frac{3}{{4}}}\frac{\left(\overline{\Delta p^{2}}\right )^{-\frac{1}{{2}}}\left (-\frac{\partial^{2} G(i,j)}{{\partial k^{2}_{i}}}\right )^{\frac{1}{{4}}}}{{\sqrt{\left (\frac{\partial^{2} G(i,j)}{{\partial k \partial c}} \right )^{2}+\left ( \frac{\partial^{2} G(i,j)}{{\partial k \partial b}}\right )^{2}}}},\nonumber \\
\label{SMxespredfin}%
\end{eqnarray}
where index $k$ is either $\alpha$ or $\beta$. This equation follows from $(\overline{x^{2}_{\epsilon}}-\overline{x_{\epsilon}}^{2})\propto DT_{k}$, taking into account $T_{k}=\sqrt{(\overline{x^{2}_{k}}-\overline{x_{k}}^{2})}/\overline{\left | v_{k} \right |}$, where $\overline{\left | v_{k} \right |}$ is the absolute average velocity of a population subjected to fluctuating payoffs $(b,c)$ near its stable point:
\begin{eqnarray}
\overline{\left | v_{k} \right |} &=& \frac{\left(\overline{\Delta p^{2}}\right )^{\frac{1}{{2}}}}{{T_{gen}}}\left ( \overline{x^{2}_{k}}-\overline{x_{i}}^{k} \right )\times\nonumber \\
&&\left |\left [ \frac{\partial}{{\partial c}}+\frac{\partial}{{\partial b}}  \right ]\frac{\partial G(\overrightarrow{r},\overrightarrow{r}')}{{\partial x_{i}}}\right |,\label{SMvabsav}
\end{eqnarray}
see eq. (\ref{SMvival1}).

The contour plot of payoff fluctuations $\overline{\Delta p^{2}}$ causing equal development rate $\overline{v_{\epsilon}}$ (\ref{SMvi2finbet}) at different points of the payoff space $(b,c)$ (with $\epsilon=0.5$) is shown in Fig. 4F.

\subsection{Information exchange vs. time}

The development rate of information exchange increases with fluctuations of payoffs $b$ and $c$, corresponding to fast fluctuations of the population size $\overline{(\Delta S/S)^{2}}$. Maximum information exchange $I$ in a population depends on the affect ability $\epsilon$, together with the payoffs $b$ and $c$. Consequently, the instantaneous development of the information exchange is:
\begin{eqnarray}
\frac{\partial I}{{\partial t}}=\frac{\partial I}{{\partial \epsilon}}\overline{v_{\epsilon}}(\epsilon,\alpha,\beta,b,c), \label{SMinfdev}
\end{eqnarray}
under assumption that the average values of $b$ and $c$ are constant. Taking into account eqs. (\ref{SMvi2finalp}), (\ref{SMvi2finbet}) and (\ref{SMxespredfin}), one can write:
\begin{eqnarray}
\frac{\partial I}{{\partial t}}\propto\frac{1}{{T_{dev}}}f(\epsilon,b,c), \label{SMinfdevav}
\end{eqnarray}
where $f$ specifies the population under discussion. The time of development of information exchange and processing $T_{dev}$:
\begin{eqnarray}
T_{dev}\propto T_{gen}(DT_{gen})^{-\frac{3}{{4}}}\left (\overline{\left (\frac{\Delta S}{{S}}\right )^2}\right )^{-\frac{1}{{2}}}, \label{SMinfotimedev}
\end{eqnarray}
is inversely proportional to the diffusion coefficient $D$ and to population size fluctuations $\sqrt{\overline{(\Delta S/S)^{2}}}$, and proportional to the time-span of a single generation $T_{gen}$.

\subsection{Derivation of eqs. (\ref{SMvival1}) and (\ref{SMvival2})}
\label{SMsecDvi12}

We are looking for a $\delta$-function like solution (\ref{SMrhodelta}) of eq. (\ref{SMrepdyn1}). In a moving frame of reference $(\overrightarrow{r}\leftarrow \overrightarrow{r}-\overrightarrow{v}t, t\leftarrow t)$, eq. (\ref{SMrepdyn1}) becomes:
\begin{equation}
-\overrightarrow{v}\frac{\partial\rho(\overrightarrow{r})}{{\partial \overrightarrow{r}}}+\frac{\partial\rho(\overrightarrow{r})}{{\partial t}}=\frac{1}{{T_{gen}}}\rho(\overrightarrow{r})\frac{F(\overrightarrow{r})-\overline F}{{|\overline F|}}+D\frac{\partial^{2}\rho}{{\partial \overrightarrow{r}^{2}}}. \label{SMrepdynvframe}%
\end{equation}
where $\overrightarrow{v}$ is the velocity of the frame of reference.

To derive the $i$th component of the velocity $v_{i}$, one should multiply eq. (\ref{SMrepdynvframe}) by the orders of $x_{i}$ (where $\overrightarrow{x}=\overrightarrow{r}-\overrightarrow{r}_{0}$) and integrate over the entire phenotype space $\overrightarrow{r}$. The equations corresponding to the orders from $x^{0}_{i}$ to $x^{3}_{i}$ are:
\begin{widetext}
\begin{eqnarray}
0 &=& \frac{1}{{T_{gen}}}\frac{F(r_{0})-\overline F}{{\overline F}}+\frac{1}{{T_{gen}\overline F}}\times\nonumber \\
&& \left (\frac{\partial G(\overrightarrow{r},\overrightarrow{r}')}{{\partial x_{k}}}\overline{x_{k}}+\frac{\partial G(\overrightarrow{r},\overrightarrow{r}')}{{\partial x'_{k}}}\overline{x'_{k}}+\right .\nonumber \\
&&\left. +\frac{1}{{2}}\frac{\partial G(\overrightarrow{r},\overrightarrow{r}')}{{\partial x_{k}\partial x_{j}}}\overline{x_{k}x_{j}}+\frac{1}{{2}}\frac{\partial G(\overrightarrow{r},\overrightarrow{r}')}{{\partial x_{k}\partial x'_{j}}}\overline{x_{k}x'_{j}}+\frac{1}{{2}}\frac{\partial G(\overrightarrow{r},\overrightarrow{r}')}{{\partial x'_{k}\partial x'_{j}}}\overline{x'_{k}x'_{j}}\right ),\nonumber \\
\label{SMdellim0}
\end{eqnarray}
\begin{eqnarray}
v_{i}+\frac{\partial \overline{x_{i}}}{{\partial t}} &=& \frac{1}{{T_{gen}}}\frac{F(r_{0})-\overline F}{{\overline F}}\overline{x_{i}}+\frac{1}{{T_{gen}\overline F}}\times \nonumber \\
&& \left (\frac{\partial G(\overrightarrow{r},\overrightarrow{r}')}{{\partial x_{k}}}\overline{x_{k}x_{i}}+\frac{\partial G(\overrightarrow{r},\overrightarrow{r}')}{{\partial x'_{k}}}\overline{x'_{k}x_{i}}+\right .\nonumber \\
&& \left. +\frac{1}{{2}}\frac{\partial G(\overrightarrow{r},\overrightarrow{r}')}{{\partial x_{k}\partial x_{j}}}\overline{x_{k}x_{j}x_{i}}+\frac{1}{{2}}\frac{\partial G(\overrightarrow{r},\overrightarrow{r}')}{{\partial x_{k}\partial x'_{j}}}\overline{x_{k}x'_{j}x_{i}}+\frac{1}{{2}}\frac{\partial G(\overrightarrow{r},\overrightarrow{r}')}{{\partial x'_{k}\partial x'_{j}}}\overline{x'_{k}x'_{j}x_{i}}\right ),\nonumber \\
\label{SMdellim1}
\end{eqnarray}
\begin{eqnarray}
2v_{i}\overline{x_{i}}+\frac{\partial \overline{x^{2}_{i}}}{{\partial t}} &=&  2D+\frac{1}{{T_{gen}}}\frac{F(r_{0})-\overline F}{{\overline F}}\overline{x^{2}_{i}}+\frac{1}{{T_{gen}\overline F}}\times \nonumber \\
&& \left (\frac{\partial G(\overrightarrow{r},\overrightarrow{r}')}{{\partial x_{k}}}\overline{x_{k}x^{2}_{i}}+\frac{\partial G(\overrightarrow{r},\overrightarrow{r}')}{{\partial x'_{k}}}\overline{x'_{k}x^{2}_{i}}+\right .\nonumber \\
&&\left .+\frac{1}{{2}}\frac{\partial G(\overrightarrow{r},\overrightarrow{r}')}{{\partial x_{k}\partial x_{j}}}\overline{x_{k}x_{j}x^{2}_{i}}+\frac{1}{{2}}
\frac{\partial G(\overrightarrow{r},\overrightarrow{r}')}{{\partial x_{k}\partial x'_{j}}}\overline{x_{k}x'_{j}x^{2}_{i}}+\frac{1}{{2}}\frac{\partial G(\overrightarrow{r},\overrightarrow{r}')}{{\partial x'_{k}\partial x'_{j}}}\overline{x'_{k}x'_{j}x^{2}_{i}}\right ),\nonumber \\
\label{SMdellim2}
\end{eqnarray}
\begin{eqnarray}
3v_{i}\overline{x^{2}_{i}}+\frac{\partial \overline{x^{3}_{i}}}{{\partial t}} &=& 6D\overline{x_{i}}+\frac{1}{{T_{gen}}}\frac{F(r_{0})-\overline F}{{\overline F}}\overline{x^{3}_{i}}+\frac{1}{{T_{gen}\overline F}}\times\nonumber \\
&& \left (\frac{\partial G(\overrightarrow{r},\overrightarrow{r}')}{{\partial x_{k}}}\overline{x_{k}x^{3}_{i}}+\frac{\partial G(\overrightarrow{r},\overrightarrow{r}')}{{\partial x'_{k}}}\overline{x'_{k}x^{3}_{i}}+\right .\nonumber \\
&&\left .+\frac{1}{{2}}\frac{\partial G(\overrightarrow{r},\overrightarrow{r}')}{{\partial x_{k}\partial x_{j}}}\overline{x_{k}x_{j}x^{3}_{i}}+\frac{1}{{2}}\frac{\partial G(\overrightarrow{r},\overrightarrow{r}')}{{\partial x_{k}\partial x'_{j}}}\overline{x_{k}x'_{j}x^{3}_{i}}+\frac{1}{{2}}\frac{\partial G(\overrightarrow{r},\overrightarrow{r}')}{{\partial x'_{k}\partial x'_{j}}}\overline{x'_{k}x'_{j}x^{3}_{i}}\right ).\nonumber \\
\label{SMdellim3}
\end{eqnarray}
\end{widetext}
The velocities of a confined population (eqs. (\ref{SMvival1}) and (\ref{SMvival2})) follow from eqs. (\ref{SMdellim0}-\ref{SMdellim3}), under assumptions:
\begin{eqnarray}
\frac{\partial \overline{x_{i}}}{{\partial t}}=\frac{\partial \overline{x^{2}_{i}}}{{\partial t}}=\frac{\partial \overline{x^{3}_{i}}}{{\partial t}}=0, \label{SMxndt0}%
\end{eqnarray}
that the population is static in the moving frame of reference.

To derive the components of the velocity $v_{i}^{1st}$ (\ref{SMvival1}) and $v_{i}^{2nd}$ (\ref{SMvival2}), one should multiply eq. (\ref{SMdellim0}) by $\overline{x_{i}}$ and subtract it from eq. (\ref{SMdellim1}). The other terms vanish, e.g $\overline{x_{k}}\overline{x_{j}}-\overline{x_{k}x_{j}}=0$ unless $k=j$.
The last term of the expression for $v_{i}^{2nd}$ (\ref{SMvival2}) is a consequence of contributions to the average fitness $\overline{F}$:
\begin{eqnarray}
\overline{F}=G(\overrightarrow{r},\overrightarrow{r})+\frac{\partial G(\overrightarrow{r},\overrightarrow{r}')}{{\partial x_{i}}}\overline{x_{i}}+\frac{\partial G(\overrightarrow{r},\overrightarrow{r}')}{{\partial x'_{i}}}\overline{x'_{i}}, \nonumber \\
\label{SMFavPder}
\end{eqnarray}
by the asymmetry of the population shape, $\overline{x_{i}},\overline{x'_{i}}\neq 0$.

\subsection{Derivation of eqs. (\ref{SMspreadsl}), (\ref{SMskewsl}), (\ref{SMspreadfa}) and (\ref{SMskewfa})}
\label{SMDspsk}

At slow propagation, the spread of a population $\left ( \overline{x^{2}_{i}}-\overline{x_{i}}^{2} \right )$ (\ref{SMspreadsl}) follows from eq. (\ref{SMdellim2}), after subtraction of eq. (\ref{SMdellim0}) multiplied by $\overline{x^{2}_{i}}$:
\begin{eqnarray}
2D=-\frac{1}{{T_{gen}}}\frac{1}{{2}}\frac{\partial^{2} G(\overrightarrow{r},\overrightarrow{r}')}{{\partial x^{2}_{i}}}\left ( \overline{x^{4}_{i}}-\overline{x_{i}^{2}}^{2} \right )
, \label{SMspread1}%
\end{eqnarray}
where the population was assumed to be near its stable position:
\begin{eqnarray}
\left (\frac{\partial G(\overrightarrow{r},\overrightarrow{r}')}{{\partial x_{i}}}\right )^{2}&\ll&\left |\frac{\partial^{2} G(\overrightarrow{r},\overrightarrow{r}')}{{\partial x^{2}_{i}}}\right |, \label{SMconfcond1}
\end{eqnarray}
\begin{eqnarray}
DT&\ll&\left |\frac{\partial^{2} G(\overrightarrow{r},\overrightarrow{r}')}{{\partial x^{2}_{i}}}\right |, \label{SMconfcond2}
\end{eqnarray}
\begin{eqnarray}
\frac{\partial^{2} G(\overrightarrow{r},\overrightarrow{r}')}{{\partial x^{2}_{i}}} &<& 0, \label{SMconfcond3}
\end{eqnarray}
and $v_{i}$ was neglected. In case of a $\delta$-function like $\rho(\overrightarrow{r})$, condition:
\begin{eqnarray}
\left ( \overline{x^{2}_{i}}-\overline{x_{i}}^{2} \right )^{2}\propto\left ( \overline{x^{4}_{i}}-\overline{x_{i}^{2}}^{2} \right ), \label{SMorderprop}%
\end{eqnarray}
is assumed to hold.

The asymmetry (\ref{SMskewsl}) is derived by substituting $D$ (eq. (\ref{SMspread1})) and $v^{1st}_{i}$ (eq. (\ref{SMvival1})) into eq. (\ref{SMdellim3}), after subtraction of (eq. (\ref{SMdellim0}) multiplied by $\overline{x^{3}_{i}}$:
\begin{eqnarray}
\overline{x_{i}} =
-\frac{\partial G(\overrightarrow{r},\overrightarrow{r}')}{{\partial x_{i}}}
\left( \frac{\partial^{2} G(\overrightarrow{r},\overrightarrow{r}')}{{\partial x^{2}_{i}}} \right )^{-1}
\frac{\left ( 3\overline{x^{2}_{i}}^{2}-\overline{x_{i}}^{4} \right )}{{\left ( -2\overline{x^{4}_{i}}+\frac{3}{{2}}\overline{x_{i}^{2}}^{2} \right )}}
,\nonumber \\
\label{SMskew1}%
\end{eqnarray}
where terms higher than $\overline{x^{4}_{i}}$ where neglected. Eq. (\ref{SMskew1}) corresponds to eq. (\ref{SMskewsl}), taking into account:
\begin{eqnarray}
\frac{\left ( 3\overline{x^{2}_{i}}^{2}-\overline{x_{i}}^{4} \right )}{{\left ( -2\overline{x^{4}_{i}}+\frac{3}{{2}}\overline{x_{i}^{2}}^{2} \right )}}\rightarrow const < 0, \label{SMskewconst}%
\end{eqnarray}
in the limit of a $\delta$-function like $\rho(\overrightarrow{r})$.

The spread (\ref{SMspreadfa}) in case of fast propagation:
\begin{eqnarray}
\left (\frac{\partial G(\overrightarrow{r},\overrightarrow{r}')}{{\partial x_{i}}}\right )^{2}&\gg&\left |\frac{\partial^{2} G(\overrightarrow{r},\overrightarrow{r}')}{{\partial x^{2}_{i}}}\right |,\nonumber \\ \label{SMfastcond1}
\end{eqnarray}
is estimated as the diffusion radius:
\begin{eqnarray}
\left ( \overline{x^{2}_{i}}-\overline{x_{i}}^{2} \right )\approx DT_{cr},\label{SMsprfa}
\end{eqnarray}
during an amount of time:
\begin{eqnarray}
T_{cr}=\frac{\sqrt{\left ( \overline{x^{2}_{i}}-\overline{x_{i}}^{2} \right )}}{{\left |v_{i}\right |}},\label{SMTcr}
\end{eqnarray}
required for a population to pass its own size. The spread as a function of $D$ and derivatives of the gain $P$ follows from (\ref{SMsprfa}), taking into account eqs. (\ref{SMvival1}) and (\ref{SMTcr}):
\begin{eqnarray}
\left ( \overline{x^{2}_{i}}-\overline{x_{i}}^{2} \right ) &\propto& \frac{D^{2}}{{v^{2}_{f}}},\nonumber \\
\label{SMspreadfa1}%
\end{eqnarray}
where velocity $v_{f}$ is:
\begin{eqnarray}
v_{f}=(T_{gen})^{-\frac{1}{{3}}}(D)^{\frac{2}{{3}}}\left (\frac{\partial G(\overrightarrow{r},\overrightarrow{r}')}{{\partial x_{f}}}\right )^{\frac{1}{{3}}}. \label{SMvf1}%
\end{eqnarray}
Asymmetry (\ref{SMskewfa}) is estimated from eq. (\ref{SMdellim2}), as $D/v_{i}$:
\begin{eqnarray}
\overline{x_{i}} &\propto&
(DT_{gen})^{\frac{1}{{3}}}\left (\frac{\partial G(\overrightarrow{r},\overrightarrow{r}')}{{\partial x_{i}}}\right )^{-\frac{1}{{3}}}, \label{SMskewfa2}%
\end{eqnarray}
taking into account (\ref{SMfastcond1}).

\subsection{Derivation of eqs. (\ref{SMmestabcondalpha}) and (\ref{SMmestabcondbeta}).}
\label{SMDabstab}

To derive eq. (\ref{SMmestabcondalpha}), consider a population composed of a host $h$ and a mutant $m$, with the phenotypes $(\epsilon_{i}^{h},\alpha_{i}^{h},\beta_{i}^{h})$ and $(\epsilon_{i}^{m},\alpha_{i}^{m},\beta_{i}^{m})$ correspondingly. The condition (\ref{SMmestabcondalpha}) follows from the stability condition (\ref{SMstabcond1}):
\begin{equation}
G(h,h)-G(m,h)>0,\label{SMPhhPmhdif}
\end{equation}
under assumption that only the host possesses a developed indirect reciprocity mode $(\alpha_{1}^{h}=1,\beta_{1}^{h}=0)$, while the mutant is different from the host by its aggression aversion $(\alpha_{1}^{m}=1-\Delta\alpha,\beta_{1}^{m}=0)$. All other behavior modes are identical:
\begin{equation}
(\epsilon_{i}^{h},\alpha_{i}^{h},\beta_{i}^{h})=(\epsilon_{i}^{m},\alpha_{i}^{m},\beta_{i}^{m}),\label{SMmodeeq} \end{equation}
for all $i \neq 1$.

The condition for evolutionary stability (\ref{SMPhhPmhdif}) becomes:
\begin{eqnarray}
\sum_{i,j}\epsilon_{i}\epsilon_{j}\left (P(h_{i}|h_{j}|W^{i}_{pq})-P(m_{i}|h_{j}|W^{i}_{pq}) \right )>0,\label{SMPhhPmhdif1}
\end{eqnarray}
taking into account the payoffs (\ref{SMgainmh}). The terms with $i\neq 1$ and $j\neq 1$ vanish due to (\ref{SMmodeeq}). The first term of (\ref{SMmestabcondalpha}) corresponds to $i=j=1$ terms of (\ref{SMPhhPmhdif1}):
\begin{eqnarray}
&&\epsilon^{2}_{1}\left ( P(\alpha=1,\beta=0|\alpha=1,\beta=0|W^{1}_{pq})-\ldots\right .\nonumber \\
&&\left .-P(\alpha=1-\Delta\alpha,\beta=0|\alpha=1,\beta=0|W^{1}_{pq}) \right )=\ldots\nonumber \\
&&=\epsilon^{2}_{1}\frac{W^{1}_{CS}-W^{1}_{SC}}{{2}},\label{SMPhhPmhdif2}
\end{eqnarray}
in the limit of $\Delta\alpha\ll1$. The difference of the payoffs in eq. (\ref{SMPhhPmhdif2}) converges to a finite value in the limit $\Delta\alpha\rightarrow 1$ as consequence of the instability of the behavior in the indirect reciprocity mode, see eqs. (\ref{SMgammalimeq}-\ref{SMW21sum}). The second term of (\ref{SMmestabcondalpha}) follows from:
\begin{eqnarray}
&&\sum_{j\neq 1}\epsilon_{1}\epsilon_{j}\left ( P(\alpha=1,\beta=0|\alpha_{j},\beta_{j}|W^{1}_{pq})-\ldots\right .\nonumber \\
&&\left .-P(\alpha=1-\Delta\alpha,\beta=0|\alpha_{j},\beta_{j}|W^{1}_{pq}) \right )\propto \epsilon_{1}\Delta\alpha.\nonumber \\
\label{SMPhhPmhdif3}
\end{eqnarray}
This sum is proportional to $\Delta\alpha$, since there are no instabilities in behavior for any $(\alpha_{j},\beta_{j})\neq (1,0)$ interacting with the modes $(\alpha=1,\beta=0)$ and $(\alpha=1-\Delta\alpha,\beta=0)$.

The condition (\ref{SMmestabcondbeta}) follows in the same way as (\ref{SMmestabcondalpha}) assuming $(\alpha_{1}^{m}=1,\beta_{1}^{m}=\Delta\beta)$, rather than the $(\alpha_{1}^{m}=1-\Delta\alpha,\beta_{1}^{m}=0)$, mutant.

\subsection{Derivation of eq. (\ref{SMWtable3}).}
\label{SMDtab}

For an average fitness (\ref{SMfitnessaverage}) $\overline{F}\approx 1$, two independent parameters are sufficient to describe the payoffs for responsive behavior in a population composed of individuals with two behavior modes. Reduction of the payoff (\ref{SMWtable1}) to a two parameter form introduces only a new time scale in the equation of motion (\ref{SMrepdyn}), without having any impact on its evolutionary dynamics.

To derive the table (\ref{SMWtable3}) from (\ref{SMWtable1}), two transformations are required:
\begin{equation}
W'_{pq}=W_{pq}-W_{SS}, \label{SMtran1}
\end{equation}
\begin{equation}
W''_{pq}=W'_{pq}/(W_{CC}-W_{SS}). \label{SMtran2}
\end{equation}
Consequently, the parameters $b$ and $c$ in (\ref{SMWtable3}) are:
\begin{eqnarray}
b&=&\frac{W_{SC}-W_{SS}}{{W_{CC}-W_{SS}}},\nonumber \\
c&=&\frac{W_{CS}-W_{SS}}{{W_{CC}-W_{SS}}}. \label{SMbc}
\end{eqnarray}
The transformation (\ref{SMtran2}) applied to eq. ((\ref{SMrepdyn})) results in:
\begin{equation}
\frac{\partial\rho(\xi)}{{\partial t}}=\frac{1}{{T_{gen}}}\rho(\xi)\frac{F(\xi|W'_{pq})-\overline F(W'_{pq})}{{|\overline F|}}+\widetilde{M}, \label{SMrepdyntran1}%
\end{equation}
This follows from:
\begin{eqnarray}
F(W_{pq})=F(W'_{pq})+W_{SS},\nonumber \\
\overline{F}(W_{pq})=\overline{F}(W'_{pq})+W_{SS}, \label{SMbc1}
\end{eqnarray}
due to the property of the payoff (\ref{SMpayoff1}):
\begin{equation}
P(m_{i}|h_{j}|W_{pq})=P(m_{i}|h_{j}|W'_{pq})+W_{SS}, \label{SMCPWtag}
\end{equation}
taking into account that $\sum_{pq}\Omega_{pq}=1$.

The second transformation (\ref{SMtran2}) brings eq. (\ref{SMrepdyntran1}) to the form:
\begin{equation}
\frac{\partial\rho(\xi)}{{\partial t'}}=\frac{1}{{T_{gen}}}\rho(\xi)\left (F(\xi|W''_{pq})-\overline F(W''_{pq})\right )+\widetilde{M}, \label{SMrepdyntran2}%
\end{equation}
where:
\begin{equation}
t'=t\frac{W_{CC}-W_{SS}}{{\bar F}}, \label{SMtimetran2}%
\end{equation}
and
\begin{equation}
W''_{pq}=\begin{tabular}{c|c|c}
  & $C$ & $S$ \\
\hline
$C$ & $1$ & $c$ \\
\hline
$S$ & $b$ & $0$  \\
\end{tabular}, \label{SMWtable4}
\end{equation}
corresponding to eq. (\ref{SMWtable3}).

\subsection{Solution of the system of eqs. (\ref{SMomegasys}).}
\label{SMdermat}

The system of eqs. (\ref{SMomegasys}) can be presented as:
\begin{equation}
\left(
  \begin{array}{ccccc}
    \alpha_{1} & 0 & \alpha_{1}-1 & 0 \\
    \alpha_{2} & \alpha_{2}-1 & 0 & 0 \\
    0 & \beta_{1} & 0 & \beta_{1}-1 \\
    0 & 0 & \beta_{2} & \beta_{2}-1 \\
    1 & 1 & 1 & 1  \\
  \end{array}
\right)
\left(
  \begin{array}{c}
    \Omega_{SS} \\
    \Omega_{SC} \\
    \Omega_{CS} \\
    \Omega_{CC} \\
  \end{array}
\right)=
\left(
  \begin{array}{c}
    0 \\
    0 \\
    0 \\
    0 \\
    1 \\
  \end{array}
\right).\label{SMomegasysmat}
\end{equation}
A solution for this system exists if the determinant of the corresponding extended matrix $U$ vanishes:
\begin{equation}
\det U = 0,\label{SMdetU0}
\end{equation}
where
\begin{equation}
U = \left(
  \begin{array}{ccccc}
    \alpha_{1} & 0 & \alpha_{1}-1 & 0 & 0 \\
    \alpha_{2} & \alpha_{2}-1 & 0 & 0 & 0 \\
    0 & \beta_{1} & 0 & \beta_{1}-1 & 0 \\
    0 & 0 & \beta_{2} & \beta_{2}-1 & 0 \\
    1 & 1 & 1 & 1 & 1 \\
  \end{array}
\right).\label{SMUmat}
\end{equation}
For a general case there is no solution, since it includes five equations and only four variables.

The determinant of $U$:
\begin{eqnarray}
&&\det U(\alpha_{1},\beta_{1},\alpha_{2},\beta_{2}) = \nonumber \\
&&\alpha_{1}(\beta_{1}-1)\beta_{2}+\alpha_{2}(\alpha_{1}\beta_{2}-\beta_{1}(\alpha_{1}+\beta_{2}-1)).\nonumber \\
\label{SMdeter1}
\end{eqnarray}
vanishes either in case of identical phenotypes $(\alpha_{1},\beta_{1})=(\alpha_{2},\beta_{2})$ or at the boundaries of the $(\alpha,\beta)$ space, $\alpha=0,1$ or $\beta=0,1$. It can be demonstrated as follows:
\begin{equation}
\det U(\alpha,\beta,\alpha,\beta) = 0,\label{SMdetUeq}
\end{equation}
Then taking into account (\ref{SMdetUeq}), $\alpha_{2}\neq\alpha_{1}$ fits (\ref{SMdetU0}) if:
\begin{equation}
\left .\frac{\partial\det U}{{\partial \alpha_{2}}}\right |_{\alpha_{1,2}=\alpha, \beta_{1,2}=\beta} = \beta(1-\beta),\label{SMdetUderalpha}
\end{equation}
vanishes. The same holds for changes in $\beta_{2}$:
\begin{equation}
\left .\frac{\partial\det U}{{\partial \beta_{2}}}\right |_{\alpha_{1,2}=\alpha, \beta_{1,2}=\beta} = -\alpha(1-\alpha).\label{SMdetUderbeta}
\end{equation}
The solution (\ref{SMomegaAB}) is valid in these cases and can be checked by a substitution.

\subsection{Derivation of eqs. (\ref{SMgammaii}) and (\ref{SMgigjigii}).}
\label{SMDgii}

In a homogenous population composed of individuals $(\epsilon,\alpha,\beta)$, the average probability $\gamma_{NR}$ of the selfish reaction $S$ is:
\begin{eqnarray}
\gamma_{NR}=\epsilon^{2}\gamma_{NR}+\epsilon(1-\epsilon)\gamma_{NR}+\ldots\nonumber \\  +(1-\epsilon)\epsilon\gamma_{R,NR}+(1-\epsilon)^{2}\gamma_{R,R}, \label{SMhomgam1}
\end{eqnarray}
where one averages over all possible interactions: non-responsive vs. non-responsive, non-responsive vs. responsive, responsive vs. non-responsive and responsive vs. responsive. The probabilities to provide reaction $S$ while interacting with non-responsive and responsive competitors are:
\begin{eqnarray}
\gamma_{R,NR}=(1-\alpha)\gamma_{NR}+(1-\beta)(1-\gamma_{NR}), \label{SMhomgam2}
\end{eqnarray}
and
\begin{eqnarray}
\gamma_{R,R}=\frac{1-\beta}{{1+\alpha-\beta}}, \label{SMhomgam3}
\end{eqnarray}
correspondingly.

Solution of eq. (\ref{SMhomgam1}) for $\gamma_{NR}$, results in eq. (\ref{SMgigjigii}).

\subsection{Derivation of eqs. (\ref{SMvfindyneps1}), (\ref{SMvfindynalp}), (\ref{SMvfindynbet}) and (\ref{SMvi2finab}).}
\label{SMDabe}

These equations are derived substituting the gain $G(i,j)$ (\ref{SMdinpay}) in expressions (\ref{SMvival1}) and (\ref{SMvi2fin}). One must take into account that:
\begin{equation}
\frac{\partial G(i|j)}{{\partial\alpha_{i}}}=\frac{\partial P(i|j)}{{\partial\alpha_{i}}}+(1-\epsilon_{i})\frac{\partial P(i|j)}{{\partial\gamma_{ji}}}\frac{\partial \gamma_{ji}}{{\partial\alpha_{i}}}, \label{SMGderalp}
\end{equation}
and
\begin{equation}
\frac{\partial^{2} G(i|j)}{{\partial\epsilon_{i}\partial\alpha_{i}}}=-\frac{\partial P(i|j)}{{\partial\gamma_{ji}}}\frac{\partial \gamma_{ji}}{{\partial\alpha_{i}}}, \label{SMGderepsalp}
\end{equation}
together with similar expressions for $\partial G(i|j)/\partial\beta_{i}$ and $\partial^{2} G(i|j)/\partial\epsilon_{i}\partial\beta_{i}$.

\subsection{Derivation of eqs. (\ref{SMalphastab}) and (\ref{SMbetastab}).}
\label{SMlast}

To derive eq. (\ref{SMalphastab}), following condition (\ref{SMstabboundpar}), one should find the points of zero velocity (\ref{SMvfindynalp}) on the $\beta=0$ axis:
\begin{equation}
\frac{\partial P(i|j)}{{\partial\alpha_{i}}}+(1-\epsilon_{i})\frac{\partial P(i|j)}{{\partial\gamma_{ji}}}\frac{\partial \gamma_{ji}}{{\partial\alpha_{j}}}=0. \label{SMvalpbound}
\end{equation}
Solving this equation for $\alpha$ results in (\ref{SMalphastab}), taking into account the expressions for payoff $P$ (\ref{SMpayoftbbc}) and $\gamma_{ji}$ (\ref{SMgammaji}).

Derivation of eq. (\ref{SMbetastab}) is analogous to the derivation of eq. (\ref{SMalphastab}). One should substitute eqs. (\ref{SMpayoftbbc}) and (\ref{SMgammaji}) into (\ref{SMvfindynbet}), finding its zeros on the $\alpha=1$ axis.

The stability condition (\ref{SMstabboundsec}) holds for the stable points (\ref{SMalphastab}) and (\ref{SMbetastab}) in cases of $(b>0,c<b)$ and $(b>1,c>b)$ correspondingly. This can be checked by substituting eqs. (\ref{SMdinpay}) and (\ref{SMpayoftbbc}) into condition (\ref{SMskew}) (matching (\ref{SMstabboundsec}) in case of a confined population):
\begin{widetext}
\begin{eqnarray}
&&\mbox{$\frac{\partial v_{\alpha}}{{\partial\alpha}}\propto-\frac{(1+b) K^3 (-1+\epsilon ) \epsilon ^3 \left(1+\left(-1+2 K^2\right) \epsilon -\sqrt{1+\left(-2+4 K^2\right) \epsilon +\epsilon ^2}\right)\left(-1+\epsilon +\sqrt{1+\left(-2+4 K^2\right) \epsilon +\epsilon ^2}\right)}{\left(-1+\epsilon -2 K \epsilon +\sqrt{1+\left(-2+4 K^2\right) \epsilon
+\epsilon ^2}\right)^2 \left(-1+\epsilon +2 K \epsilon +\sqrt{1+\left(-2+4 K^2\right) \epsilon +\epsilon ^2}\right)^3},$}\nonumber \\
&&\mbox{$K=\frac{c}{{b}},$}\label{SMstabsecalp}
\end{eqnarray}
\end{widetext}
and
\begin{widetext}
\begin{eqnarray}
&&\mbox{$\frac{\partial v_{\beta}}{{\partial\beta}}\propto\frac{b (-1+\epsilon ) \epsilon ^3 \left(K (-1+\epsilon )+\sqrt{K^2 (-1+\epsilon )^2+4 \epsilon }\right) \left(K^2 (-1+\epsilon )-2 \epsilon+K \sqrt{K^2 (-1+\epsilon )^2+4 \epsilon }\right)}{\left(K (-1+\epsilon )-2 \epsilon +\sqrt{K^2 (-1+\epsilon )^2+4 \epsilon }\right)^2 \left(K (-1+\epsilon)+2 \epsilon +\sqrt{K^2 (-1+\epsilon )^2+4 \epsilon }\right)^3},$}\nonumber \\
&&\mbox{$K=\frac{c-1}{{b-1}}.$}\label{SMstabsecbet}
\end{eqnarray}
\end{widetext}
These expressions are negative in the regions $(b>0,c<b)$ and $(b>1,c>b)$ correspondingly.

The condition (\ref{SMstabboundper}) holds for the stable points (\ref{SMalphastab}) and (\ref{SMbetastab}) in the cases of $(b>0,c<b)$ and $(b>1,c>b)$ correspondingly, assuming equal probabilities to be the second to respond (see (\ref{SMomegaeqi})) in the course of an interaction. It can be checked by substituting the general expression for the payoffs $P$ (\ref{SMpayoff1}) together with eq. (\ref{SMomegaeqi}) in eqs. (\ref{SMdinpay}), (\ref{SMvfindynalp}) and (\ref{SMvfindynbet}). The derived expression for $v_{\perp}$ are too cumbersome to be present here, though they can be handled with the help of a symbolic calculation software, e.g. Wolfram Mathematica.

\end{document}